\documentclass[10pt,letterpaper]{article}
\usepackage{amsmath, amssymb, amsfonts, graphics}
\usepackage[numbers]{natbib}
\usepackage[sectionbib]{chapterbib}
\usepackage[top=1in, bottom=.75in, left=.75in, right=.75in]{geometry}
\usepackage{titling}

\begin{document}

\bibliographystyle{osajnl}

\title{\Large{Quantitative, Comparable Coherent Anti-Stokes Raman Scattering (CARS) Spectroscopy: Correcting Errors in Phase Retrieval}}
\author{\large{Charles H. Camp Jr.\thanks{To whom correspondances should be address} , Young Jong Lee, Marcus T. Cicerone}\\\\
\large{\textit{Biosystems and Biomaterials Division, National Institute of Standards and Technology,}}\\
\large{\textit{100 Bureau Dr, Gaithersburg, MD 20899, USA}}\\\\
\large{charles.camp@nist.gov}}
\date{\vspace{-5ex}}

\maketitle
\begin{abstract}
Coherent anti-Stokes Raman scattering (CARS) microspectroscopy has demonstrated significant potential for biological and materials imaging. To date, however, the primary mechanism of disseminating CARS spectroscopic information is through pseudocolor imagery, which explicitly neglects a vast majority of the hyperspectral data. Furthermore, current paradigms in CARS spectral processing do not lend themselves to quantitative sample-to-sample comparability. The primary limitation stems from the need to accurately measure the so-called nonresonant background (NRB) that is used to extract the chemically-sensitive Raman information from the raw spectra. Measurement of the NRB on a pixel-by-pixel basis is a nontrivial task; thus, reference NRB from glass or water are typically utilized, resulting in error between the actual and estimated amplitude and phase. In this manuscript, we present a new methodology for extracting the Raman spectral features that significantly suppresses these errors through phase detrending and scaling. Classic methods of error-correction, such as baseline detrending, are demonstrated to be inaccurate and to simply mask the underlying errors. The theoretical justification is presented by re-developing the theory of phase retrieval via the Kramers-Kronig relation, and we demonstrate that these results are also applicable to maximum entropy method-based phase retrieval. This new error-correction approach is experimentally applied to glycerol spectra and tissue images, demonstrating marked consistency between spectra obtained using different NRB estimates, and between spectra obtained on different instruments. Additionally, in order to facilitate implementation of these approaches, we have made many of the tools described herein available free for download.
\end{abstract}

\section*{Introduction}
Coherent anti-Stokes Raman scattering (CARS) is a nonlinear scattering phenomenon in which two photons, ``pump" and ``Stokes", coherently excite a molecular vibration. From the excited mode, a ``probe" photon inelastically scatters off with an energy increase equal to that of the vibrational state. This optical technique affords label-free, molecularly-sensitive investigation of samples without autofluorescence competition and at significantly higher speeds than offered by traditional spontaneous Raman spectroscopy\cite{Zumbusch1999,Petrov2007}. CARS microscopies and microspectroscopies have demonstrated success on a large variety of material and biological samples ranging from polymer blends\cite{Kee2004,Vacano2007} to brain tumor masses\cite{Evans2007,CampJr2014}, requiring fractions of a second for a micrograph highlighting a single vibrational mode to a few minutes for a complete hyperspectral image.

A fundamental challenge to harnessing the information content in CARS microspectroscopies, is extraction of the chemically-specific Raman signal from the so-called ``nonresonant background" (NRB). The NRB is predominantly comprised of electronic signal contributions from other nonlinear optical phenomena that are less chemically specific\cite{Muller2007}. Although it is sometimes viewed as an interference, the NRB actually amplifies the weak Raman signal, enabling high-sensitivity detection\cite{Muller2007,CampJr2014}. This coherent mixing, however, does bring spectral distortions. For CARS microscopies that probe small increments of the full vibrational spectrum, physical methods are utilized to reduce the NRB generation, which in turn reduces the overall CARS signal \cite{Cheng2001b,Ganikhanov2006,Dudovich2002,Muller2007}. For spectroscopic CARS techniques (microspectroscopies), however, two classes of numerical methods are commonly used to remove the distortion of the NRB: one based on maximizing entropy \cite{Vartiainen1992} and the other utilizing the Kramers-Kronig (KK) relation \cite{Liu2009}. These techniques are functionally equivalent \cite{Cicerone2012} and have demonstrated success with materials\cite{Lee2011,Baldacchini2009} and biological samples\cite{Rinia2007a,Rinia2006,Masia2013,Parekh2010,CampJr2014}. Both of these techniques, however, rely upon an accurate measurement of the NRB, which also incorporates the system excitation profile and spectral response. To date, no approach has been found to conveniently and accurately measure the NRB, so retrieved spectra generally have phase and amplitude errors. Additionally as typically implemented, these techniques implicitly assume a constancy of the nonresonant background at every pixel across the image, which is typically a poor assumption for heterogeneous samples. The resulting processed spectra contain baseline fluctuations, distorted peaks, and provide limited quantifiable information between molecular species; thus, intra- and inter-sample comparisons are significantly hampered. Additionally, the measurements performed on different CARS instruments-- or even the same instrument under altered operating conditions-- are not directly comparable.

In this work, we present a new method of processing CARS spectra that suppresses errors resulting from the use of an inexact reference NRB spectra, removing baseline fluctuations and generating spectra that are agnostic to the reference material used. In upcoming sections, we present a theoretical re-examination of the KK relation and errors resultant from reference NRB measurements. Use of the KK presents analytical expressions for correcting these errors, but these results are also broadly applicable to the maximum entropy method (MEM)\cite{Vartiainen1992}. Finally, we present simulated and experimental results using neat glyercol and murine pancreas tissues. Spectra are presented from two broadband CARS (BCARS) architectures with dramatically different excitation profiles\cite{Parekh2010,Lee2010,CampJr2014}. The underlying purpose of this manuscript is to enable the extraction of ``pre-processed" spectra that are universally comparable in amplitude and shape. This facilitates dissemination of pre-processed spectra as a new ``currency" of coherent Raman imaging data.

\section{Theory}
Classically, the CARS spectral intensity, $I_{CARS}$, may be described as\cite{Gomez1996}:
\begin{align}
I_{CARS}(\omega) ~\propto~ \left | \iiint \chi^{(3)}(\omega)E_{p}(\omega_p)E^*_{S}(\omega_S)E_{pr}(\omega_{pr})\times \delta \left (\omega - \omega_p + \omega_S - \omega_{pr})d\omega_p d\omega_S d\omega_{pr}\right)\right |^2,
\end{align}
where $\chi^{(3)}$ is the nonlinear susceptibility of the sample; $E_{p}$, $E_{S}$, and $E_{pr}$ are the pump, Stokes, and probe field amplitudes, within the frequency spaces, $\omega_p$, $\omega_S$, and $\omega_{pr}$, respectively; and the delta function ensures energy conservation. This equation may be re-written in a more intelligible form\cite{CampJr2014}:
\begin{align}
I_{CARS}(\omega) ~\propto&~ \left | \left \lbrace \underbrace{\left [E_S(\omega)\star E_p(\omega)\right ]}_{C_{st}(\omega)}\chi^{(3)}(\omega) \right \rbrace\ast E_{pr}(\omega)\right |^2\\
\nonumber\\
~\approxeq&~ |\widetilde{C}_{st}(\omega)|^2|\widetilde{\chi}^{(3)}(\omega)|^2,
\end{align}
where $\star$ and $\ast$ are the cross-correlation and convolution operations, respectively, and $C_{st}$ is the coherent stimulation profile. Equations 1 and 2 are mathematically equivalent. If we assume a spectrally narrow probe source, we can introduce an ``effective" stimulation profile, $\widetilde{C}_{st}$, and nonlinear susceptibility, $\widetilde{\chi}^{(3)}$, as presented in eq 3, where $\widetilde{C}_{st}(\omega)\equiv \left [C_{st}(\omega)\ast E_{pr}(\omega)\right ]/\int E_{pr}(\omega) d\omega$ and $\widetilde{\chi}^{(3)}(\omega) \equiv \chi^{(3)}(\omega)\ast E_{pr}(\omega)$.

The nonlinear susceptibility describes signal contributions from Raman vibrationally-resonant, $\chi_R$, and vibrationally non-resonant, $\chi_{NR}$, sources: $\chi^{(3)}(\omega) = \chi_R(\omega) + \chi_{NR}(\omega)$. To a first degree approximation\cite{Tolles1977}, spontaneous Raman spectra, $I_{Raman}$, are related to the vibrationally resonant component of the CARS spectra as: $I_{Raman}(\omega) \propto \text{Im}\lbrace \chi_R(\omega) \rbrace$, where `Im' indicates the imaginary component. The purpose of phase retrieval is to ascertain a phase, $\phi(\omega)$, that isolates the Raman resonant components from the total nonlinear susceptibility.
 
\subsection{Phase Retrieval Using the Kramers-Kronig Relation}
The KK relation states that there is an explicit relationship between the real and imaginary components of a function, $f(\omega)$, that is causal (analytic\cite{Toll1956,Lucarini2005}); thus, if only the real (or imaginary) part is known, the imaginary (or real) part can be calculated. In CARS and other spectroscopies, neither the real nor imaginary portion of $\chi^{(3)}$ is accessible (\textit{n.b.}: $\widetilde{C}_{st}$ in eq 3 is not a causal function). If the function is square-integrable, there also exists an explicit relationship between the complex norm of the function and the phase\cite{Smith1977}:
\begin{align}
&\ln|f(\omega)| ~=~ -\hat{\mathcal{H}}\{\phi(\omega)\}\\
&\phi(\omega) ~=~ \hat{\mathcal{H}}\{\ln|f(\omega)|\},
\end{align}
where $\hat{\mathcal{H}}$ is the Hilbert transform\cite{Poularikas1999}, which assumes knowledge of the complex modulus of the function or phase over an infinite frequency range.  Practically, however, the CARS spectral recording window is limited; thus, we will introduce a ``windowed" Hilbert transform: $\hat{\mathcal{H}}_W$, as:
\begin{align}
&\hat{\mathcal{H}}_W\lbrace f(x);\omega_a,\omega_b \rbrace  ~=~ \frac{\mathcal{P}}{\pi}\int_{\omega_a}^{\omega_b} \frac{f(x')}{x-x'}dx'\\
&\lim_{
\begin{subarray}{l}
\omega_a \rightarrow -\infty\\\omega_b \rightarrow \infty
\end{subarray}}
\hat{\mathcal{H}}_W\lbrace f(x);\omega_a,\omega_b \rbrace ~=~ \hat{\mathcal{H}}\lbrace f(x) \rbrace,
\end{align}
which is limited to the spectral range $\omega_a$ to $\omega_b$ (for compactness, these parameters will be neglected from the operator form). $\mathcal{P}$ is the Cauchy principle value. Under the conditions that (a) the Raman peaks encompassed within this window are not affected by those outside of the window and (b) any resonances of $\chi_{NR}$ are far-removed from those of $\chi_R$ (as is typically the case with infrared stimulation), the windowed- and analytic Hilbert transform are related as:
\begin{align}
\hat{\mathcal{H}}_W\left\lbrace \frac{1}{2}\ln |\widetilde{\chi}^{(3)}(\omega)|^2\right\rbrace ~\approx~ \hat{\mathcal{H}}\left\lbrace \frac{1}{2}\ln |\widetilde{\chi}^{(3)}(\omega)|^2\right\rbrace + \epsilon(\omega),
\end{align}
where $\epsilon(\omega)$ is an additive error term (see Figure S1 and associated Supporting Information demonstrating the additive nature). Applying eqs 5 and 8 to eq 3, the retrieved phase from the raw CARS spectrum, $\phi_{CARS}$, may be described as:
\begin{align}
\phi_{CARS}(\omega) ~=~ \hat{\mathcal{H}}_W\left\lbrace \frac{1}{2}\ln I_{CARS}(\omega) \right\rbrace ~\approx&~ \epsilon(\omega) + \hat{\mathcal{H}}_W\left\lbrace \frac{1}{2}\ln \left |\widetilde{C}_{st}(\omega)\right|^2\right\rbrace +\hat{\mathcal{H}}\left\lbrace \frac{1}{2}\ln \left | \widetilde{\chi}^{(3)}(\omega)\right |^2 \right \rbrace\\
~=&~ \epsilon(\omega) + \hat{\mathcal{H}}_W\left\lbrace \frac{1}{2}\ln |\widetilde{C}_{st}(\omega)|^2\right\rbrace+ \angle\left[\chi_R(\omega) + \chi_{NR}(\omega)\right],
\end{align}
where $\angle$ denotes the angle (phase). The retrieved phase is not simply that of the nonlinear susceptibility but also contains contributions from the windowing error and the effective stimulation profile. If one can measure the NRB spectrum without Raman components, $I_{NRB}$, and assuming that the spectrum is far-removed from electronic resonances such that $\chi_{NR}$ is approximately real, then the following phase retrieval can be utilized in lieu of eq 10:
\begin{align}
\phi_{CARS/NRB}(\omega) ~=~ \hat{\mathcal{H}}_W\left\lbrace \frac{1}{2}\ln \frac{I_{CARS}(\omega)}{I_{NRB}(\omega)} \right\rbrace ~\approx&~ \epsilon(\omega) + \hat{\mathcal{H}}_W\left\lbrace \frac{1}{2}\ln |\widetilde{C}_{st}(\omega)|^2\right\rbrace\nonumber\\
~-&\left[~\epsilon(\omega) + \hat{\mathcal{H}}_W\left\lbrace \frac{1}{2}\ln |\widetilde{C}_{st}(\omega)|^2\right\rbrace\right]\nonumber\\
~+&~ \angle\left[\chi_{R}(\omega)+\chi_{NR}(\omega)\right]-\angle\chi_{NR}(\omega)\nonumber\\
~\approx&~\angle\left[\chi_{R}(\omega)+\chi_{NR}(\omega)\right],
\end{align}
which is analogous to applying the KK relation to $I_{CARS}/I_{NRB}$. Using this ratio as our signal, the retrieved complex spectra, $I_{retr}$, is:
\begin{align}
&I_{retr}(\omega) ~= \sqrt{\frac{I_{CARS}(\omega)}{I_{NRB}(\omega)}}\exp i \phi_{CARS/NRB} ~\approx~ \frac{|\widetilde{\chi}^{(3)}(\omega)|}{|\widetilde{\chi}_{NR}(\omega)|}\exp i \angle\left[\chi_{R}(\omega)+\chi_{NR}(\omega)\right],
\end{align}
and the Raman-like spectrum, $I_{RL}$:
\begin{align}
&I_{RL}(\omega) ~= \text{Im}\left\lbrace I_{retr}(\omega)\right\rbrace\approx \frac{\text{Im}\left\lbrace \chi_R(\omega)\right\rbrace}{|\chi_{NR}|};
\end{align}
thus, the Raman-like spectrum is proportional to the spontaneous Raman spectrum scaled by the nonresonant component. The earliest KK relation results\cite{Liu2009} (presenting a very different derivation) utilized $\sqrt{I_{CARS}}\sin \phi_{CARS/NRB}$, which is directly proportional to the spontaneous Raman spectrum, but it implicitly assumes that $\widetilde{C}_{st}(\omega)$ is constant and does not account for $\epsilon(\omega)$. Later work\cite{Masia2013}, using an identical phase retrieval method (although with a different derivation-- see the Supporting Information), identified a need to remove the stimulation profile via $I_{CARS}/I_{NRB}$. Together these works explain the method of phase retrieval under certain ideal conditions. In the following sections, we will present the ramifications of the real-world condition when the NRB of the sample is not directly measurable. This analysis, as only enabled by the derivations in eqs 8 through 13, will lead to a direct method to account and correct for these errors.

\subsection{Errors from Inaccurate NRB Measurement}
Although phase-retrieval is theoretically straightforward, measuring the NRB is technically challenging. To circumvent this limitation, researchers rely on measurements from model materials, such as coverslip glass or water. This practice assumes that the model material does not contain a significant Raman signature and the NRB varies little from material to material. Additionally in practice, any reference material spectral deviations from the actual NRB are assumed to contribute to a slowly-varying baseline that can be subtracted off via various detrending methods. With the increased sensitivity of spectroscopic CARS systems, however, these assumptions appear more and more naive. As demonstrated below, the difference between the NRB and reference measurement does not lead to an additive error, but rather a multiplicative complex error.

We obtain a reference measurement, $I_{ref}$, as a surrogate for a proper NRB measurement. Here, $I_{ref}(\omega)=\xi(\omega)I_{NRB}(\omega)$, and $\xi(\omega)$ is assumed to be real and positive. By applying eq 11:
\begin{align}
\phi&_{CARS/ref}(\omega) = \phi_{CARS/NRB}(\omega) + \underbrace{\hat{\mathcal{H}}_W\left\lbrace \frac{1}{2}\ln  \frac{1}{\xi(\omega)}\right\rbrace}_{\phi_{err}(\omega)},
\end{align}
we see that the Raman-like spectrum (eq 13), $I_{RL-ref}$, is:
\begin{align}
I_{RL-ref}(\omega) &~=~ \underbrace{\sqrt{\frac{1}{\xi(\omega)}}}_{A_{err}(\omega)}\sqrt{\frac{I_{CARS}(\omega)}{I_{NRB}(\omega)}}\sin\left[\phi_{CARS/NRB}(\omega) + \phi_{err}(\omega)\right]
\end{align}
From eqs 14 and 15, one can see that the use of a reference measurement leads to both amplitude ($A_{err}$) and phase ($\phi_{err}$) distortions. Accordingly, removing these errors is not simply a matter of subtraction. The phase error, however, is additive in nature, and connected to the amplitude error via the KK relation:
\begin{align}
\phi_{err}(\omega) ~=&~ \hat{\mathcal{H}}\left\lbrace \ln A_{err}(\omega)\right\rbrace\\
\ln A_{err}(\omega) ~=&~ -\hat{\mathcal{H}}\left\lbrace \phi_{err}(\omega)\right\rbrace.
\end{align}
There is, however, an ambiguity in this relationship. If $\xi(\omega)$ is multiplied by a constant, $\Xi$: $\phi_{err}(\omega) = \hat{\mathcal{H}}\left\lbrace \ln 1/\Xi\xi(\omega)\right\rbrace = \hat{\mathcal{H}}\left\lbrace \ln 1/\xi(\omega)\right\rbrace$, since the Hilbert transform of a constant is zero.

\subsection{Correcting Phase Error and Scale}
The purpose of correcting for Raman signature extraction errors is not simply to generate qualitatively accurate spectra, but those that are quantitatively reliable, facilitating direct comparison and analysis of spectra collected of different samples with potentially different reference materials, and on various spectroscopic systems having different excitation profiles. In the previous subsection we demonstrated that the use of a reference NRB which only approximates the nonresonant response of the material induces amplitude and phase distortions. Additionally, the commonly used method of subtracting baseline fluctuations, borrowed from the spontaneous Raman community, does not actually remove the errors as (1) the nature of the error is complex valued and (2) the amplitude error is multiplicative.

Below we show that one can properly correct for signal extraction error using the following steps:
\begin{enumerate}
\item Remove phase error via phase detrending, and correct for part of the amplitude error via the KK relation
\item Correct for scaling error (involving $\Xi$) and the error from the windowed Hilbert transform (of $\phi_{err}$) via unity-centering of the real component of the retrieved (phase-corrected) spectrum.
\end{enumerate}
As displayed in eq 14, the difference between the ideal phase retrieval (in which the NRB of the sample is exactly known) and that using a model material is $\phi_{err}$, which is additive. The retrieved phase (ideal) is qualitative similar to Raman-like spectra in that the spectral features are peaks that extend positively from a zero baseline. A slowly-varying phase error will cause a slowly-varying deviation from the zero baseline. Finding $\phi_{err}$; therefore, is a matter of isolating the erroneous baseline, for which one may use traditional baseline detrending methods (albeit, applied to the phase rather than the Raman-like spectral amplitude). From this extracted $\phi_{err}$, using eq 17, one can find the amplitude error. With these variables in hand, one can multiply the retrieved complex spectrum by a complex phase-correction multiplier, generating a phase-corrected spectrum, $I_{pc}$:
\begin{align}
I_{pc}&(\omega) ~=~ \sqrt{\frac{I_{CARS}(\omega)}{I_{ref}(\omega)}}\exp \left [i \phi_{CARS/ref}(\omega) \right ] \left\lbrace\frac{1}{\exp \left [-\hat{\mathcal{H}}\left \lbrace \phi_{err}(\omega)\right \rbrace\right]} \exp \left [-i \phi_{err}(\omega)\right]\right\rbrace
\end{align}
\textit{n.b.}: the use of ``phase-corrected" in this context is unrelated to that in Ref. \citenum{Masia2013}. As described in the previous subsection, however, there is an ambiguity in the KK relation between the error phase and amplitude. Thus the calculated $A_{err}$ is only accurate to within a constant multiplier. Additionally, the Hilbert transform in eq 18 is actually a windowed Hilbert transform; thus, $\hat{\mathcal{H}}\left \lbrace \phi_{err}(\omega)\right \rbrace = \hat{\mathcal{H}}_W\left \lbrace \phi_{err}(\omega)\right \rbrace + \epsilon_{err}(\omega)$, where $\epsilon_{err}$ is a window-effect error term similar to that introduced in eq 8.

To finalize the error correction, one needs to account for the $A_{err}$ ambiguity and $\epsilon_{err}$. Both of these quantities are discoverable by analyzing the real component of the phase-corrected spectrum in eq 18 since the real component of eq 12 is unity-centered, i.e., $\left \langle |\widetilde{\chi}^{(3)}|/|\widetilde{\chi}_{NR}| \cos \phi_{CARS/NRB}\right \rangle = 1$. The existence of $\Xi$, however, will alter the mean; thus, one could calculate the mean of the real component of the retrieved spectrum and normalize the real and imaginary components by this value. $\epsilon_{err}$, however, may impart a frequency-dependent component to this mean. Using numerical means, though, one can find a slowly-varying center-line and normalize the phase-corrected spectrum; thus, removing $\Xi$ and $\epsilon_{err}$ in one step. Assuming this centerline can be found, a complex corrected spectrum $I_{corrected}$ may be calculated:
\begin{align}
I_{corrected}(\omega) ~=&~ \frac{I_{pc}(\omega)}{\left \langle \text{Re}\lbrace I_{pc}(\omega) \rbrace \right \rangle(\omega)}\\
~=&~ \frac{|\widetilde{\chi}^{(3)}(\omega)|}{|\widetilde{\chi}_{NR}(\omega)|}\exp i \phi_{CARS/NRB}(\omega),
\end{align}
where we have noted the frequency-dependence of the mean-line in the denominator of eq 19. Comparison of eqs 12 and 20 shows that using the prescribed steps, one can retrieve the same spectrum using a reference NRB as if the NRB were measurable.

\section{Materials and Methods}
\subsection{CARS Microspectroscopy}
In this manuscript, most CARS spectra were collected on a recently introduced broadband CARS (BCARS) system that has been described elsewhere\cite{CampJr2014}. In order to demonstrate that properly retrieved spectra can be essentially identical, irrespective of instrumentation, some spectra were collected on an instrument that uses an earlier implementation of BCARS signal generation \cite{Parekh2010,Lee2011}. The newer BCARS system excites molecular vibrations more efficiently with the highest response at the lowest wavenumbers, whereas, the traditional BCARS system excites most Raman transitions with relatively uniform response. For the newer BCARS system, the total average incident power was $<$25 mW (3.5 ms integration time), and for the traditional BCARS system $<$60 mW (7.8 ms integration time).

\subsection{CARS Simulations}
The CARS simulation software was developed in MATLAB (Mathworks) and numerically implements eq 2 directly. All excitation sources were simulated as real Gaussian functions, and the Raman response a complex Lorentzian (damped harmonic oscillator) as:
\begin{align}
\chi_R(\omega) = \sum_m \frac{A_m}{\Omega_m - \omega - i\Gamma_m},
\end{align}
where $A_m$, $\Omega_m$, and $\Gamma_m$ describe the amplitude [multiplier], wavenumber, and half-width of the $m^{th}$ Raman peak. Specific parameters for the simulations are provided below.

\subsection{Signal Pre-Processing}
Image and spectral processing are performed using in-house software developed in MATLAB. Specific details of the pre-processing of BCARS spectra (or images) are presented in the Supporting Information (see Figures S2 - S11 and Table S1). In brief, dark spectra are collected as are NRB spectra from reference materials (glass coverslip, glass microscope slide, or water). For noise reduction of BCARS hyperspectral data, we utilize singular value decomposition (SVD)\cite{Cicerone2012a} on Anscombe transformed spectra. The Anscombe transform\cite{Seymour1998,Makitalo2013} normalizes the noise variance, accounting for mixed Poisson-Gaussian noise. Pertinent singular values are selected by noise analysis in the spectral and spatial domains in an automated or semi-automated fashion (Figures S8, S9). Once SVD is performed, the variance-stabilized, noise-reduced spectra are returned to their normal mixed-noise state using an optimized, generalized inverse Anscombe transformation\cite{Makitalo2013}.

Raman-like spectra are retrieved using the Hilbert transform implementation of the KK relation (described later). The erroneous component of the retrieved phase is found in an automated fashion using an asymmetric lease-squares (ALS) technique with a Whittaker smoother\cite{Eilers2005,Eilers2003,Urbas2011}. Phase and partial amplitude correction are performed as described in eq 18. To determine the mean trend line for final spectral correction (eq 19), a Savitky-Golay filter is utilized.

\subsubsection{Phase Retrieval Using the Hilbert Transform}
The Hilbert transform (eq 6) is implemented in the time-domain ($t$):
\begin{align}
\hat{\mathcal{H}}_W\left \lbrace f(\omega) \right \rbrace = \mathcal{F}\left\{i~ \text{sgn}(t)~\mathcal{F}^{-1}\left\{f(\omega)\right\}\right\},
\end{align}
where $\mathcal{F}$ and $\mathcal{F}^{-1}$ are the Fourier and inverse Fourier transforms, respectively, $\text{sgn}(t)$ is the signum (``sign") function and $f(\omega)$ is a spectrally dependent function (e.g., $I_{CARS}/I_{ref}$). Additionally, we implement a spectral padding procedure \cite{Vartiainen1992} to extend the window range, reducing numerical edge effects. This method efficiently retrieves phase with only two Fourier transforms and was designed for parallel processing. 100 parallel solutions, for example, with each spectrum containing 1000 spectral points requires $\sim$200 $\mu s$ per spectrum on a personal computer (16 GB RAM, 3.4 GHz quad-core processor). The KK and associated Hilbert transform code is freely available at http://github.com/CoherentRamanNIST in the MATLAB and Python languages.

\section{Results}
\subsection{Simulated Spectra}
\begin{figure}[!ht]
\begin{center}
\includegraphics{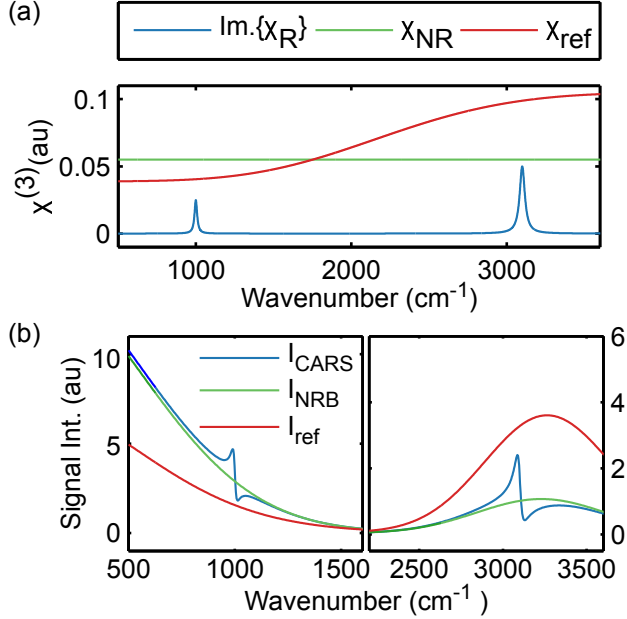}
\caption{Simulated nonlinear susceptibilities and CARS spectra. (a) Raman ($\chi_{R}$) and nonresonant nonlinear susceptibilities of the sample ($\chi_{NR}$) and of a reference ($\chi_{ref}$). (b) Simulated CARS spectra of the sample ($I_{CARS}$, $\chi = \chi_R + \chi_{NR}$), the sample in absence of Raman components ($I_{NRB}$), and of the reference material ($I_{ref}$). }
\label{SimRaw}
\end{center}
\end{figure}

To validate the presented theory on phase retrieval and error correction, we begin with the simplified case of a two-peak Raman system with parameters (eq 21): $A_1$ = 0.25, $\Omega_1$ = 1000 cm$^{-1}$, $\Gamma_1$ = 10 cm$^{-1}$, $A_2$ = 1, $\Omega_2$ = 3100 cm$^{-1}$, and $\Gamma_2$ = 20 cm$^{-1}$). $\chi_{NR}$ = 0.55 and $\chi_{ref}$ is $\chi_{NR}$ multiplied by a Gaussian function. The simulated nonlinear susceptibilities are presented in Figure \ref{SimRaw} a. The CARS spectra that result from these susceptibilities are shown in Figure \ref{SimRaw} b.

\begin{figure}[!ht]
\begin{center}
\includegraphics{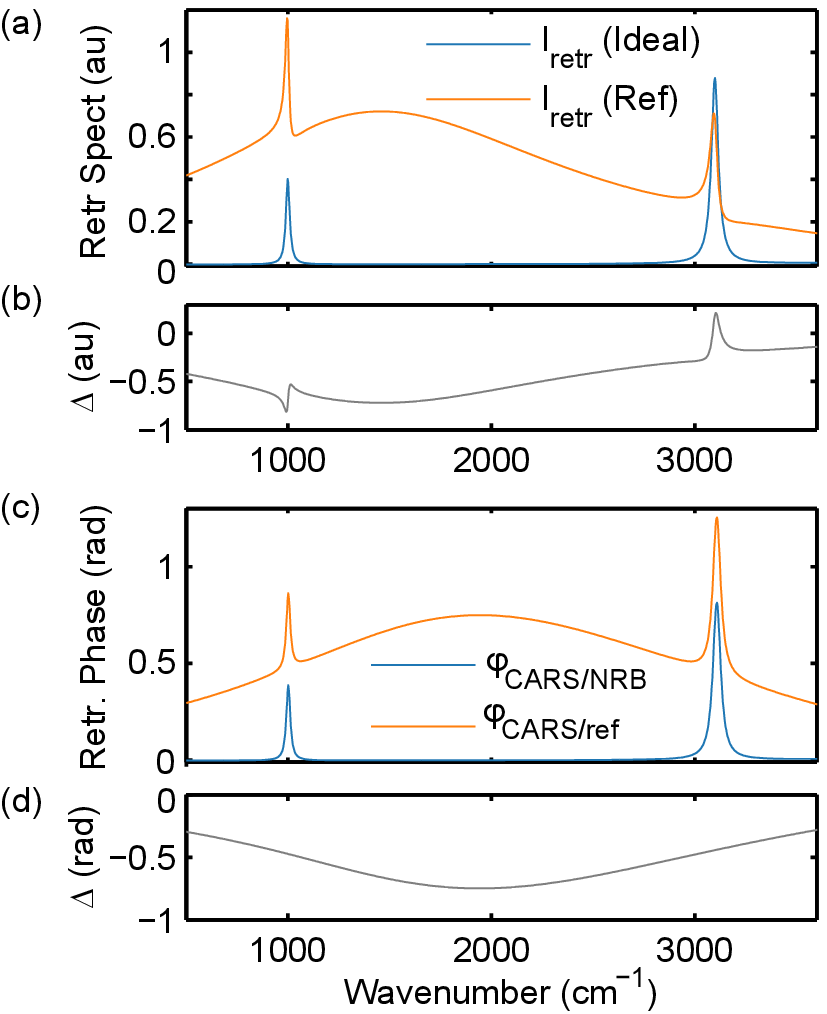}
\caption{Comparison of retrieved spectra using the KK under ideal conditions and with use of am NRB reference. (a) Retrieved spectrum ($I_{retr}$) when the NRB is exactly known (Ideal) and when a reference is utilized (Ref). (b) The difference, $\Delta$, between the ideal and reference retrieval showing remnants of the original Raman peaks. (c) The KK-retrieved phase when the NRB is known ($\phi_{CARS/NRB}$) and when using a reference ($\phi_{CARS/ref}$). The phase difference, $\Delta$, between the ideal and reference retrieval that shows only a smooth baseline with no Raman peak information.}
\label{SimPhaseAmp}
\end{center}
\end{figure}

Figure \ref{SimPhaseAmp} a shows the retrieved spectra using a reference or the actual NRB (``ideal"). The reference-retrieved spectrum show a clear, large baseline and distorted (asymmetric) peak shapes. Additionally, the peak amplitudes are sufficiently perturbed that the 1000 cm$^{-1}$ peak appears $\sim$50\% larger than the 3100 cm$^{-1}$, although the latter should be the larger of the two. Figure \ref{SimPhaseAmp} b shows the difference ($\Delta$) between the ideal and nonideal retrieved spectra. From this, one can gather that the traditional tactic of baseline detrending will only resolve the slowly-varying baseline, but the underlying peak-errors (amplitude and phase) will remain. Figure \ref{SimPhaseAmp} c shows the phase retrieved under ideal and nonideal conditions, which does not display any obvious peak distortions. As clearly presented from the difference of the retrieved phases (Figure \ref{SimPhaseAmp} d), there is no spectral distortion of the Raman peaks, but only the slowly-varying baseline ($\phi_{err}$ in eq 14).

\begin{figure}[!ht]
\begin{center}
\includegraphics{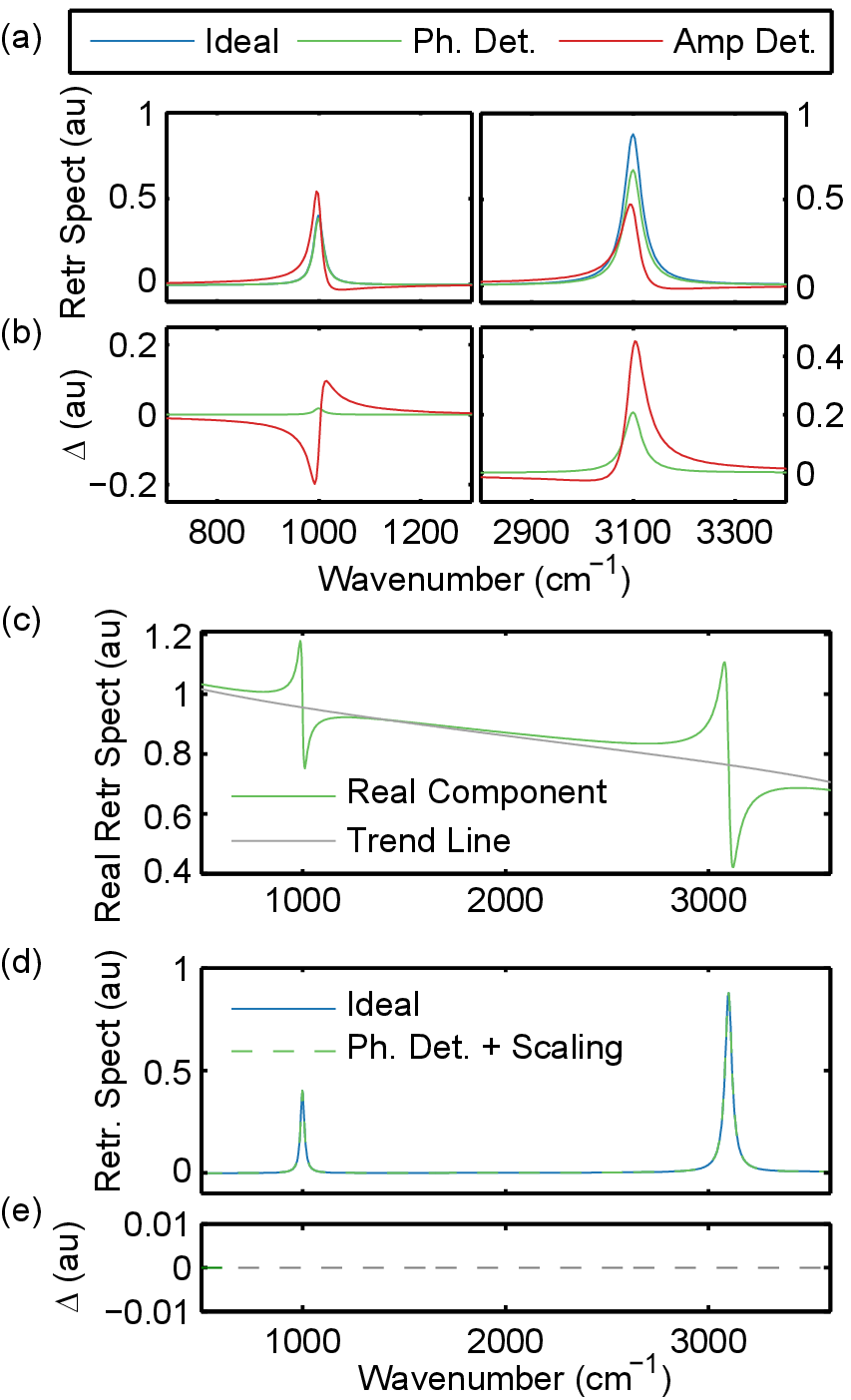}
\caption{Residual error and their correction using the proposed phase-detrending and scaling with comparison to tradition amplitude detrending. (a) KK-retrieved spectra when the NRB is known (Ideal) and using a reference material with phase detrending (Ph. Det., see eq 18) and amplitude detrending (Amp. Det.). (b) The difference, $\Delta$, between the ideal and corrected spectra. Amplitude detrending creates asymmetric peak distortion not present with phase detrending. (c) Using the mean-trend of the real component of the retrieved spectrum allows for correction of edge effects and scaling ambiguity. (d) Phase-detrended and scaled spectrum (Ph. Det. + Scaling, imaginary portion of eq 19) is identical to the ideal retrieval (Ideal), with (e) no difference, $\Delta$.}
\label{SimDeTScale}
\end{center}
\end{figure}

Using this phase error and applying a calculated amplitude correction (eq 18), the baseline and asymmetric spectral distortions are removed entirely, as shown in Figure \ref{SimDeTScale} a (differences shown in  Figure \ref{SimDeTScale} b). For reference, a traditional amplitude detrending is also displayed, showing the remaining distortions clearly. Although the phase-corrected spectrum is qualitatively similar to the ideal, the relative amplitudes of the two peaks are still incorrect, owing to $\epsilon_{err}$ and the amplitude ambiguity described previously. Figure \ref{SimDeTScale} c shows the real part of the phase-corrected spectrum and the mean trend-line that deviates from unity. Using this trend as a scaling factor (eq 19), Figure \ref{SimDeTScale} d demonstrates that the phase-corrected spectrum is now identical to the ideal retrieval (difference shown in Figure \ref{SimDeTScale} e).

As presented elsewhere\cite{Cicerone2012}, the KK and MEM phase retrieval methods are functionally equivalent. Figure S12 demonstrates the applicability of the presented phase error correction method to Raman-like spectra extracted from simulated BCARS spectra via the MEM method. Like the KK demonstration above, the prescribed method enables a reference NRB spectrum to be utilized and generates a corrected Raman-like spectrum that is equivalent to one extracted when the NRB is exactly known.

\subsection{Experimentally Measured Spectra}
\begin{figure}[!ht]
\begin{center}
\includegraphics{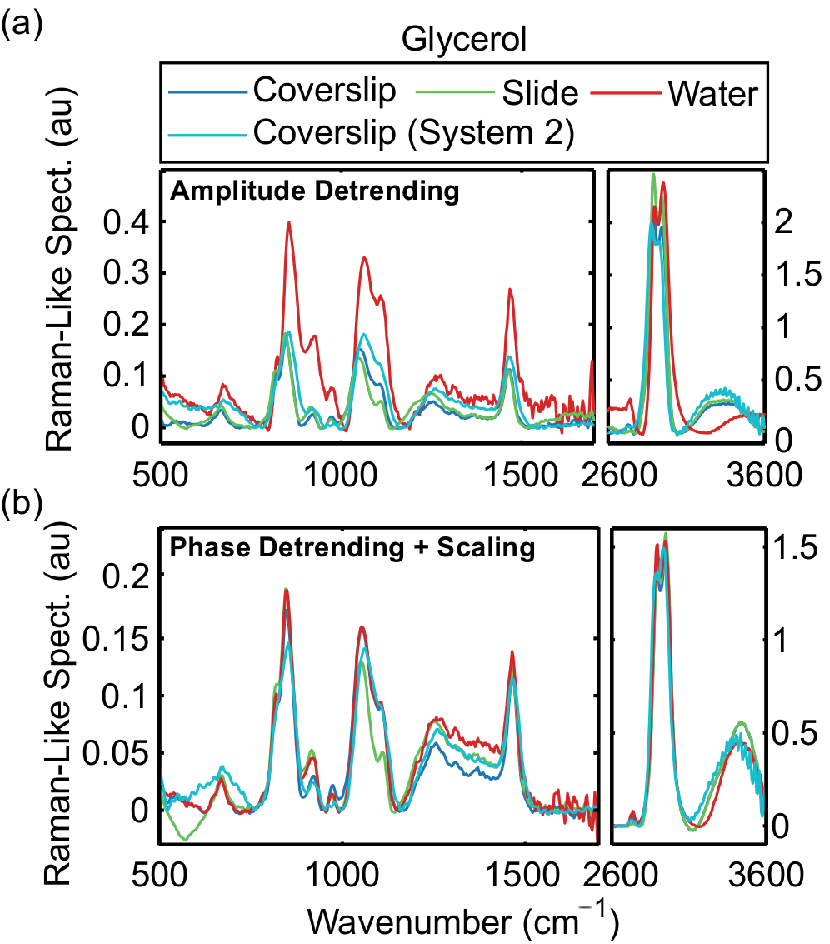}
\caption{Spectral correction of neat glycerol. (a) KK-retrieved Raman-like spectra using NRB reference materials with traditional amplitude-detrending, highlighting the inability of baseline detrending to remove amplitude and phase error. (b) Corrected spectra with phase detrending and scaling (imaginary portion of eq 19), showing close agreement with residual differences primarily arising from reference material Raman peaks.}
\label{Glycerol}
\end{center}
\end{figure}

The developed error correction method can readily be applied to experimental results without modification. Additionally, this method provides spectra that are comparable between microscopy platforms. Figure S13 a shows CARS spectra collected for neat glycerol on two BCARS systems (average of 100 acquisitions), demonstrating widely different system responses for the same molecule. Figure S13 b shows the recorded reference spectra for 3 different commonly used model materials. These reference spectra not only demonstrate a great variety of overall amplitude but also spectral features. As expected, retrieving the Raman-like spectra using these references produces amplitude and phase errors, resulting in distortions as shown in Figure S13 c. Figure \ref{Glycerol} a shows the Raman-like spectra with the slowly-varying baseline removed. The spectrum retrieved using water demonstrates the most severe distortions. Even the spectra using glasses (slide and coverslip) demonstrate differences. In comparison, Figure \ref{Glycerol} b shows the same four spectra after full correction (eq 19). The four spectra are significantly more similar in amplitude and shape. Additionally, one should notice the partial recovery of the OH-stretch peaks ($\sim$3300 cm$^{-1}$) for the spectrum retrieved using water. When a particular reference material is utilized, the retrieved spectra will have suppressed peaks wherever the model material contains Raman peaks. Within spectroscopic CARS literature, coverslip or slide glass have often served as a convenient reference. What was not apparent at the time, however, is that these glasses have nontrivial, glass-dependent Raman peaks. The primary cause of the spectral differences in Figure \ref{Glycerol} b are due to these reference material Raman peaks. Figure S15 b shows the retrieved (and corrected) spectra of these reference materials, with their peaks exactly correlating with the deviations in the retrieved spectra (Figure S15 a). These spectra were collected using a time-windowing, self-referencing method (see Supporting Information and Figure S14). This enables collecting an NRB-dominant spectrum directly from the sample, which can then be used as the reference. This also enables the use of reference material spectra with their Raman peaks suppressed. Figure S15 c shows Raman-like (corrected) spectra of glycerol using different references with their peaks suppressed.

\subsection{Tissue Imaging}
\begin{figure*}[!ht]
\begin{center}
\includegraphics{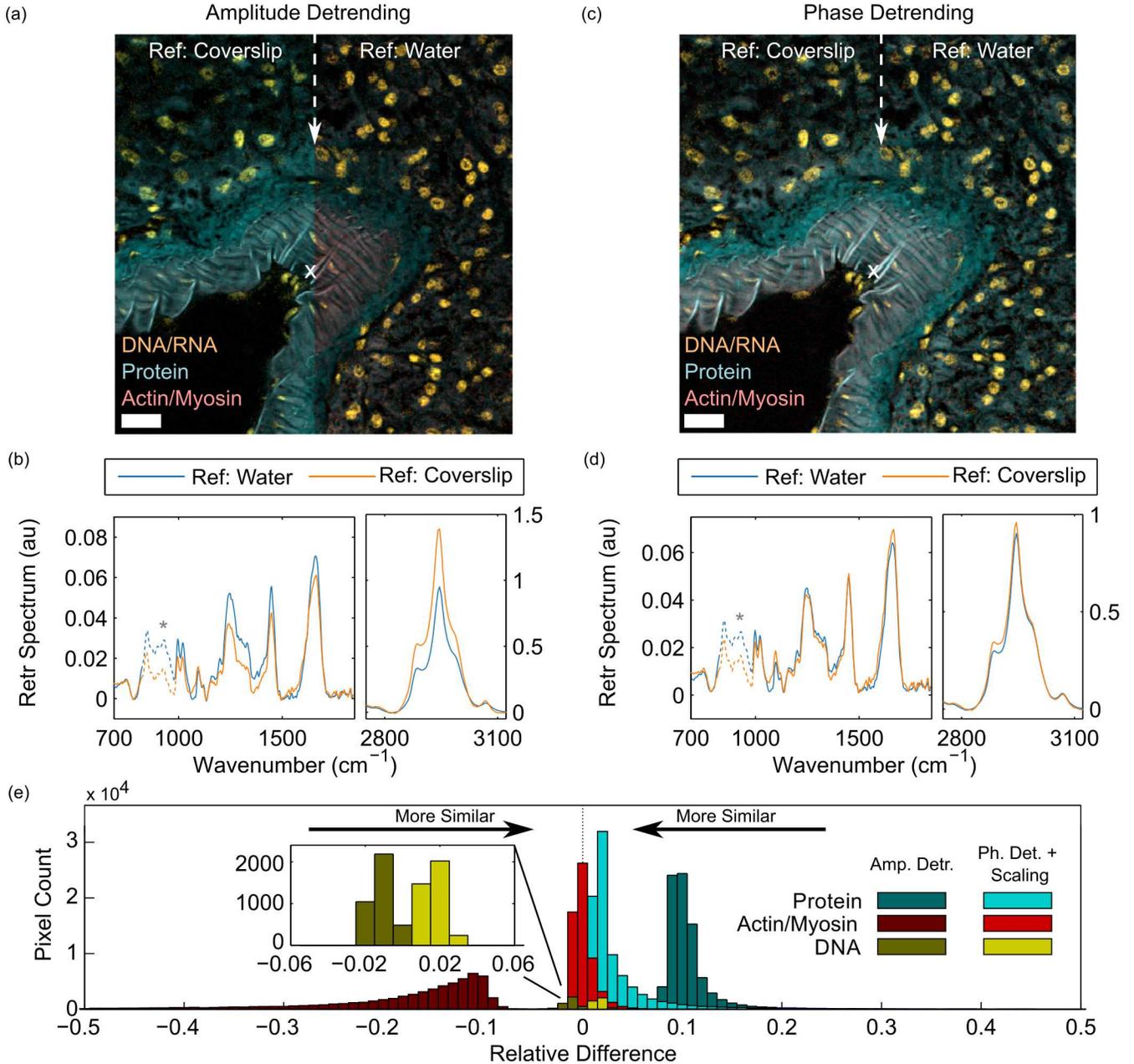}
\caption{Spectral retrieval and correction of hyperspectral data: murine pancreatic artery tissue section. (a) Traditional method of preparing pseudocolor imagery shows distinct differences when using a glass coverslip or water as the NRB reference, which cannot be simply corrected with normalization. (b) Single-pixel spectra (white 'x' in (a)) show intra-spectral deviations that are reference-dependent.  Additional errors due to Raman peaks emanating from the reference NRB materials (`*' and dashed lines). (c) Pseudocolor imagery using the presented phase retrieval shows no obvious sign of difference between the halves processed with different NRB references. (d) Single-pixel spectra show close agreement with residual error due to Raman peaks emanating from the reference NRB materials (`*' and dashed lines). (e) Histogram analysis of the relative difference between spectral peak intensities used in the creation of (a). The developed method shows $<$4\% difference (mean) between peak amplitudes.}
\label{Pancreas}
\end{center}
\end{figure*}

The presented phase retrieval method can also be reliably applied to hyperspectral images. For this purpose, we imaged a histological section of murine pancreatic artery. A 200 $\mu m$ x 200 $\mu m$ section (90,000 pixels total) was imaged with 3.5 ms dwell times. Reference spectra were collected from water and the sample coverslip. The raw BCARS image was de-noised using SVD on Anscombe-stabilized spectra, keeping 23 singular values. After this de-noising, the hyperspectral data were processed four times: twice with each reference spectrum and twice with amplitude or phase detrending methods. Figures \ref{Pancreas} (a,c) are the pseudocolor images of the murine skin highlighting protein in blue (2937 cm$^{-1}$- 2882 cm$^{-1}$), DNA in orange (785 cm$^{-1}$), and in red a shoulder peak that is dominant in the smooth muscle (1339 cm$^{-1}$) and is tentatively assigned to actin/myosin\cite{Romer1998,Buschman2001}. The Supporting Information provides more detail regarding how the images were processed. The left-halves of Figures \ref{Pancreas} (a,c) show the processing performed using the coverslip NRB reference, and the right-halves using water. The color intensity range of red, green, and blue channels are the same for both half-images within a single figure; although, the range is different between Figure \ref{Pancreas} a and Figure \ref{Pancreas} c. With amplitude detrending (Figure \ref{Pancreas} a), the boundary between the half-images is obvious. Using the coverslip reference, the blue channel (protein) is more intense than when using water. Conversely, the red channel (actin/myosin) is suppressed when using coverslip. The DNA/RNA signature is nearly the same in both images. This highlights that distortions vary across the spectrum, and one cannot simply normalize out the errors. Using phase detrending (and scaling) in Figure \ref{Pancreas} b, there is no obvious discontinuity between the two half-images. Figure 5b shows single-pixel spectra collected from a portion of the internal elastic lamina (marked by a white 'x' in Figures \ref{Pancreas} a) and corrected using amplitude detrending. In these spectra the coverslip-processed spectrum is $\sim$40\% stronger at higher energies, but $\sim$15\% to 30\% weaker within the fingerprint region. With phase detrending, shown in Figure \ref{Pancreas} d, the peak differences are significantly less obvious, with the predominant cause of residual error being from Raman peaks inherent to the reference materials (in Figures \ref{Pancreas} (b,c) significant perturbations induced by reference NRB Raman peaks in coverslip glass are denoted with a `*' and dashed lines). Figure S16 shows a histogram comparison of the Raman peak amplitudes whether performing amplitude detrending or phase detrending (and scaling), demonstrating significantly increased similarity when using the developed method. To quantify the improvement, we calculated on a pixel-by-pixel basis the relative difference (Figure \ref{Pancreas} e) between the peaks used to construct Figures \ref{Pancreas} (a,c), i.e., the coverslip-processed intensity minus the water-processed intensity, over the mean. Using amplitude detrending, the mean relative difference is 10.5\% (standard deviation, $\sigma$, 3.7\%) for protein, -17.4\% ($\sigma$=8.4\%) for actin/myosin, and -1.1\% ($\sigma$ = 0.5\%) for DNA/RNA. Using phase detrending and scaling, the mean relative difference for protein improves to 3.5\% ($\sigma$ = 4.5\%) and to 0.08\% ($\sigma$ = 1.1\%) for actin/myosin. The mean relative difference for DNA/RNA remains nearly the same (in amplitude) at 1.7\% ($\sigma$ = 0.5\%). 

\section{Summary and Discussion}
 The aim of this work is to foster quantitative reliability and repeatability of CARS spectra and to promote quantitative analysis of hyperspectral data cubes rather than rely on the pseudocolor imagery. Although this work presents the first efforts towards making CARS spectra reliable, repeatable, and universally comparable, there are still many avenues to improve upon this method. For example, the retrieved spectra using this (and other) methods are normalized to the NRB. As described in eq 12, this results from the practical necessity of removing the system response, which may contain spectral fluctuations due to laser source spectral profiles or filter set characteristics. One solution may be a thoroughly characterized reference material of which the resonant and nonresonant nonlinear susceptibilities are known; thus, enabling the extraction of the system response. This extracted system response could then be applied to the analyses of future CARS acquisitions on that system. Another opportunity is the examination of the proposed method under the condition that the nonresonant nonlinear susceptibility is not approximately real, but rather complex. The ramifications of a complex NRB may become more appreciable due to water absorption as CARS systems move further into the infrared or due to multiphoton electronic absorption as more exotic fluorescent species such as plants or algal cells are investigated.
 
 In conclusion, this work presents an expanded method of Raman signal extraction that corrects for amplitude and phase errors that are ubiquitous in CARS microspectroscopy. The presented technique, does not require any augmentation to current acquisition work flows and performs the corrections \textit{in silico}. The corrected spectra, as demonstrated with neat liquids and tissues images, significantly reduces the intra-spectral distortions caused by the use of NRB reference spectra, and facilitates direct, quantitative comparison between samples and microscopy systems. In the future, it is the hope of the authors that this method (and future improvements) will enable mass dissemination of coherent Raman hyperspectral data cubes for community data mining and analysis.

\subsection*{Corresponding Author}
E-mail: charles.camp@nist.gov

\subsection*{Disclaimer}
Any mention of commercial products or services is for experimental clarity and does not signify an endorsement or recommendation by the National Institute of Standards and Technology. The authors declare no competing financial interests.

\bibliography{QuantBCARS}
\clearpage
\bibliographystyle{osajnl}
\setcounter{equation}{0}
\setcounter{figure}{0}
\setcounter{table}{0}
\setcounter{page}{1}
\setcounter{section}{0}

\makeatletter
\renewcommand{\theequation}{S\arabic{equation}}
\renewcommand{\thefigure}{S\arabic{figure}}
\renewcommand{\thesection}{S\arabic{section}}
\renewcommand{\bibnumfmt}[1]{[S#1]}
\renewcommand{\citenumfont}[1]{S#1}

\title{\Large{Supplementary Information}}
\author{\large{Charles H. Camp Jr.$^*$, Young Jong Lee, Marcus T. Cicerone}\\\\
\large{\textit{Biosystems and Biomaterials Division, National Institute of Standards and Technology,}}\\
\large{\textit{100 Bureau Dr, Gaithersburg, MD 20899, USA}}\\\\
\large{charles.camp@nist.gov}}
\date{\vspace{-5ex}}

\maketitle

\section{Theory}
\subsection{Additive Nature of Window-Edge Effects with the Numerical Hilbert Transform}
Within the main text of the manuscript it was proposed that the full Hilbert transform and the ``windowed" Hilbert transform were related by an additive error term, $\epsilon$, under the condition that the resonant component of the nonlinear susceptibility, $\chi_R$, was fully captured and the nonresonant component, $\chi_{NR}$, was not. That is:
\begin{align}
\hat{\mathcal{H}}\left \{ \log |\chi_R(\omega) + \chi_{NR}(\omega)|\right\} ~=&~ \hat{\mathcal{H}}_W\left \{ \log |\chi_R(\omega) + \chi_{NR}(\omega)|\right\} + \epsilon(\omega) \\
~=&~ \angle\left[ \chi_R(\omega) + \chi_{NR}(\omega)\right],
\end{align}
where $\angle$ is the phase.
\begin{figure}[!ht]
\begin{center}
\includegraphics{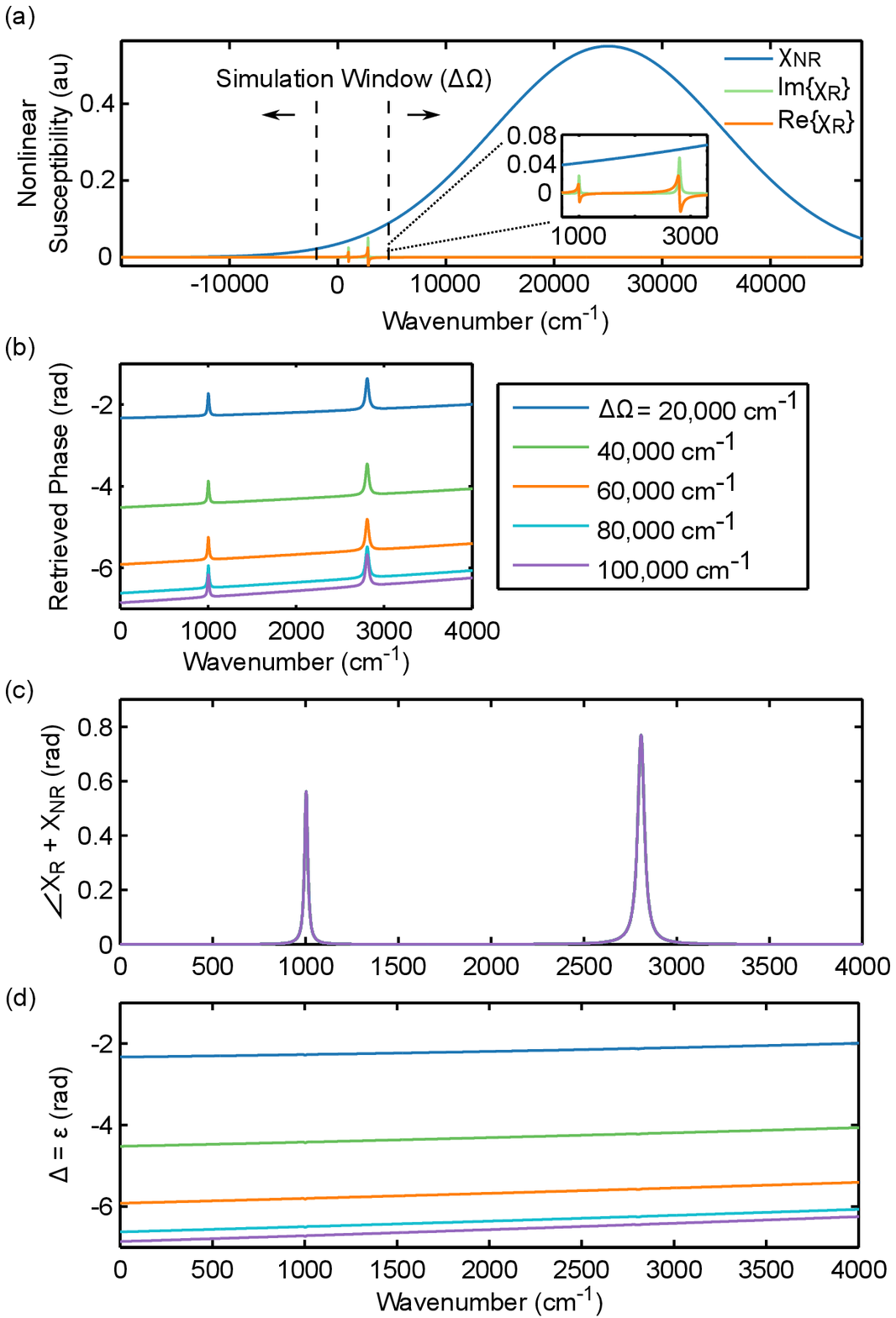}
\caption{Window edge effect produces linear, additive error. (a) Simulated $\chi_R$ and $\chi_{NR}$. The simulation window is broad enough to always capture the Raman peaks. (b) Phase retrieved via the windowed Hilbert transform over varying spectral windows, showing good agreement in peak amplitudes and shapes but with varying baselines. (c) Exact phase of $\chi_R + \chi_{NR}$. (d) Difference of the retrieved (b) and exact (c) phases, showing only a smooth baseline. This indicates that the window edge effect is  (or approximately) additive.}
\label{NonlinearSusceptWindowEdge}
\end{center}
\end{figure}

To demonstrate this relationship, we simulated a nonlinear susceptibility with two Raman peaks and an extremely broad (25,000 cm$^{-1}$ full-width half-max) nonresonant term as shown in Figure \ref{NonlinearSusceptWindowEdge} a. The windowed Hilbert transform of the complex norm of the nonlinear susceptibility was performed over larger-and-larger windows and compared with the ideal solution (the phase). Figure \ref{NonlinearSusceptWindowEdge} b shows the retrieved phase via the windowed Hilbert transform with varying window widths. Figure \ref{NonlinearSusceptWindowEdge} c shows the numerically-calculated phase of the nonlinear susceptibility, which is the same regardless of window width. The difference between the retrieved phase and the exact phase is shown in Figure \ref{NonlinearSusceptWindowEdge} d. As seen there are no remnants of Raman peaks; thus, this remainder is additive and represents the error term, $\epsilon$.

\subsection{Comparison of Previous Derivations of the Kramers-Kronig Relation to Coherent Raman Spectra}
Previous use of the KK\cite{Liu2009,Masia2013} utilized amplitude step-functions in the time-domain to effectively perform the Hilbert transform. Below we will show that these methods are, in fact, equivalent in phase retrieval and both perform a Hilbert transform. These derivations have been altered to conform to the notation used within this manuscript and a particular Fourier transform convention. Accordingly, the derivations we present may have different multiplicative constants, but the fundamental results are identical to the original derivations.

\subsubsection{The ``Time-Domain Kramers-Kronig Transform" (TDKK)}
Liu, Lee, and Cicerone\cite{Liu2009} developed a time-domain Kramers-Kronig transform with a mathematical relaxation of causality. Chiefly, the Fourier-transformed (time-domain) CARS spectrum is cut at $t < 0$ and the time-domain NRB spectra is conversely cut at $t \geq 0$. Explicitly:
\begin{align}
\phi(\omega) ~=&~ -2 \text{Im}\left\{\psi\{\ln[ I_{CARS}(\omega),I_{NRB}(\omega) ]\} - \frac{\ln I_{CARS}(\omega)}{2} \right\},
\end{align}
where `Im' selects the imaginary component, and $\psi$ is an operator defined as:
\begin{align}
\psi\{\ln[ I_{CARS}(\omega),I_{NRB}(\omega) ]\} ~\triangleq&~ \mathcal{F}\left\{u(t)\mathcal{F}^{-1}\{\ln I_{CARS}(\omega)\} + u(-t)\mathcal{F}^{-1}\{\ln I_{NRB}(\omega)\}\right\}.
\end{align}
In eq 4, $u(t)$ is a step-function defined as 1 for $t \geq 0$. The operator $\psi$ selects the Fourier transform of the CARS signal for $t\geq0$ and the Fourier transform of the NRB signal for $t<0$. Combining eqs 3 and 4 and noting that $\mathcal{F}\{u(t)\} = \sqrt{\pi/2}[\mathcal{P}/(i\pi\omega) + \delta(\omega)]$, where $\mathcal{P}$ is the Cauchy principle value:
\begin{align}
\phi(\omega) ~=&~ -\frac{2}{\sqrt{2\pi}}\Im\left\{\left[\frac{\mathcal{P}}{i\sqrt{2\pi}\omega} + \sqrt{\frac{\pi}{2}}\delta(\omega)\right]\ast \ln I_{CARS}(\omega)\right.+\nonumber\\
&\left.\left[\frac{\mathcal{P}}{i\sqrt{2\pi}(-\omega)} + \sqrt{\frac{\pi}{2}}\delta(-\omega)\right]\ast \ln I_{NRB}(\omega)- \frac{\ln I_{CARS}(\omega)}{2}\right\}\nonumber\\
~=&~ \frac{\mathcal{P}}{\pi\omega}\ast \ln I_{CARS}(\omega)-\frac{\mathcal{P}}{\pi\omega}\ast \ln I_{NRB}(\omega),
\end{align}
where $\ast$ is the convolution operation. Using the definition of the Hilbert transform for an arbitrary function $f(x)$:
\begin{align}
\hat{\mathcal{H}}\{f(x)\} = \frac{\mathcal{P}}{\pi}\int_{-\infty}^{\infty} \frac{f(x')}{x - x'}dx' = f(x)\ast \frac{\mathcal{P}}{\pi x},
\end{align}
and combining with eq 5:
\begin{align}
\phi(\omega) =& \hat{\mathcal{H}}\{ln I_{CARS}(\omega)\} - \hat{\mathcal{H}}\{\ln I_{NRB}(\omega)\}= \hat{\mathcal{H}}\left\{\ln\frac{I_{CARS}(\omega)}{I_{NRB}(\omega)}\right\} =  \hat{\mathcal{H}}\left\{\frac{|\widetilde{\chi}^{(3)}(\omega)|}{|\widetilde{\chi}_{NR}(\omega)|}\right\};
\end{align}
thus, the retrieved phase is identical with that presented in the main text of the manuscript.

\subsubsection{The ``Phase-Corrected Kramers-Kronig" (PCKK)}
Masia, \textit{et al.}\cite{Masia2013} presented a phase retrieval method in which prior to the Kramers-Kronig transform the CARS signal is normalized by the NRB reference spectrum. Afterwards a step-function is applied in the time-domain and a Fourier-transform applied:
\begin{align}
\phi(\omega) ~=&~ -2\Im \left\{\mathcal{F}\left\{u(t)\mathcal{F}^{-1}\left\{\ln \frac{I_{CARS}(\omega)}{I_{NRB}(\omega)}\right\}\right\}\right\}\nonumber\\
~=&~ -\frac{2}{\sqrt{2\pi}}\Im\left\{\left[\frac{\mathcal{P}}{i\sqrt{2\pi}\omega} + \sqrt{\frac{\pi}{2}}\delta(\omega)\right]\ast \ln\frac{I_{CARS}(\omega)}{I_{NRB}(\omega)}\right\}
\end{align}
Applying the definition of the Hilbert transform, eq 6, to eq 8:
\begin{align}
\phi(\omega) ~=&~ \frac{\mathcal{P}}{\pi\omega}\ast \ln\frac{I_{CARS}(\omega)}{I_{NRB}(\omega)} = \hat{\mathcal{H}}\left\{\ln\frac{I_{CARS}(\omega)}{I_{NRB}(\omega)}\right\} =  \hat{\mathcal{H}}\left\{\frac{|\widetilde{\chi}^{(3)}(\omega)|}{|\widetilde{\chi}_{NR}(\omega)|}\right\}.
\end{align}
From a phase-retrieval point-of-view, the results between the TDKK and the PCKK are identical (eqs 7 and 9) and with the Hilbert transform derivation presented within the main text. In application, there is a difference between the PCKK and the TDKK: amplitude normalization by the NRB. In the TDKK, the Raman-like spectrum is:
\begin{align}
I_{TDKK}(\omega) = \sqrt{{I_{CARS}(\omega)}}\sin \phi = |\widetilde{C}_{st}(\omega)||\widetilde{\chi}^{(3)}(\omega)|\sin \phi,
\end{align}
where $\widetilde{C}_{st}$ is the effective stimulation profile presented within the main text of this manuscript. The TDKK manuscript neglected the shape of the excitation sources; thus, the retrieved spectrum was directly proportion the spontaneous Raman spectrum. The PCKK manuscript accounted for these sources as:
\begin{align}
I_{PCKK}(\omega) = \sqrt{\frac{I_{CARS}(\omega)}{I_{NRB}(\omega)}}\sin \phi = \frac{|\widetilde{\chi}^{(3)}(\omega)|}{|\widetilde{\chi}_{NR}(\omega)|}\sin \phi;
\end{align}
thus, the stimulation profile is removed but the output spectrum is now scaled with respect to the NRB.

Under an ideal circumstance, the stimulation profile would be directly measurable, removed, and the retrieved spectrum would follow that of the TDKK. In practice, however, this is not trivial. Normalization by the NRB signal, as presented in the PCKK and this manuscript, removes the stimulation profile and other static, spectral perturbations such as the optical filter passband oscillations. Thus, this practice is a near-necessity for creating spectra that are directly comparable from system-to-system. As addressed within the Discussion section of the main manuscript, this may be alleviated with future development of reference materials or other techniques of quantifying the optical system spectral profile.

\section{Materials and Methods}
\subsection{Processing Methodology and Performance}
\begin{figure}[!ht]
\begin{center}
\includegraphics{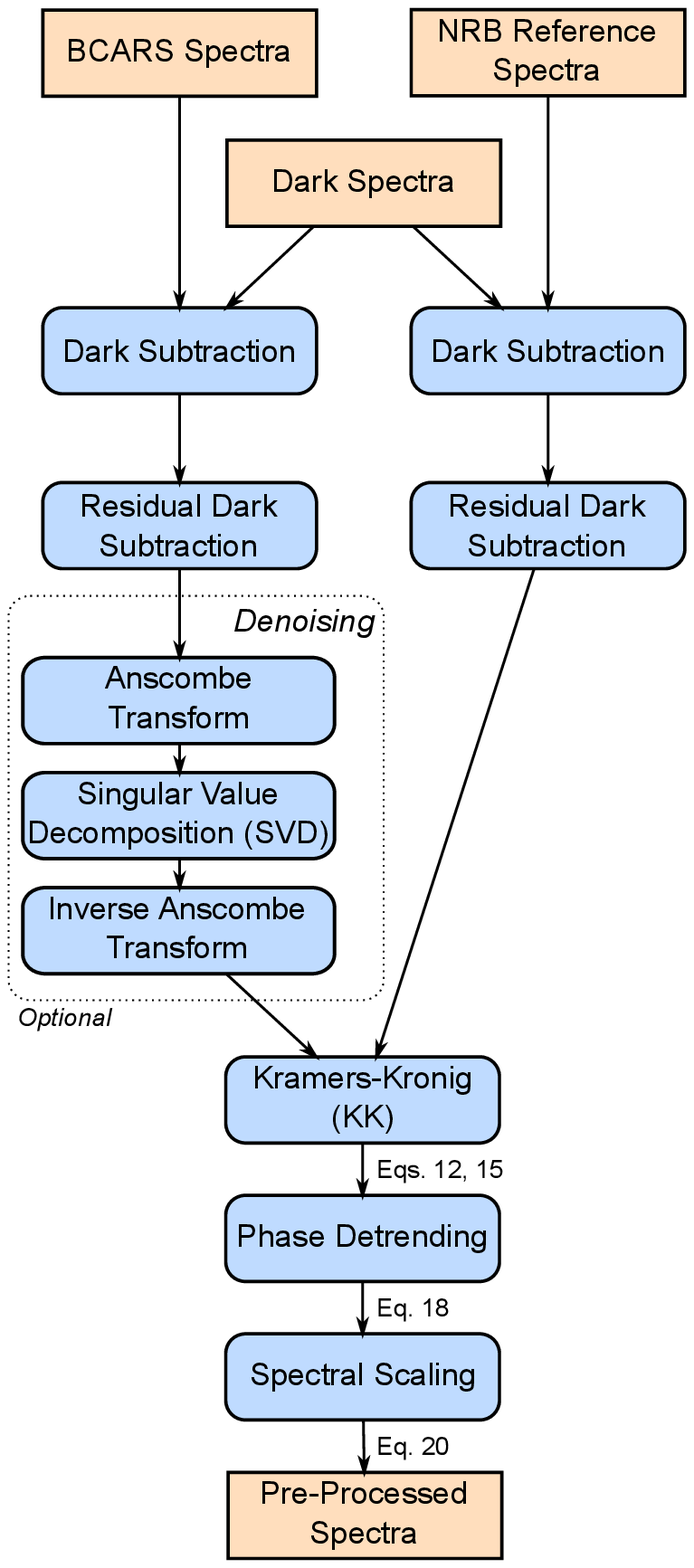}
\caption{Flow chart detailing the method of extracting quantitatively reliable Raman-like spectra from BCARS. Results of certain operations have callouts to their corresponding equations within the main text.}
\label{FlowChart}
\end{center}
\end{figure}

The typical steps taken in order to extract the Raman features from BCARS spectra or images are outlined in Figure \ref{FlowChart}. Some of these steps are common, others less so. As such, we will detail some of these steps in further detail below.

\subsubsection{Residual Background Subtraction}
\begin{figure}[!ht]
\begin{center}
\includegraphics{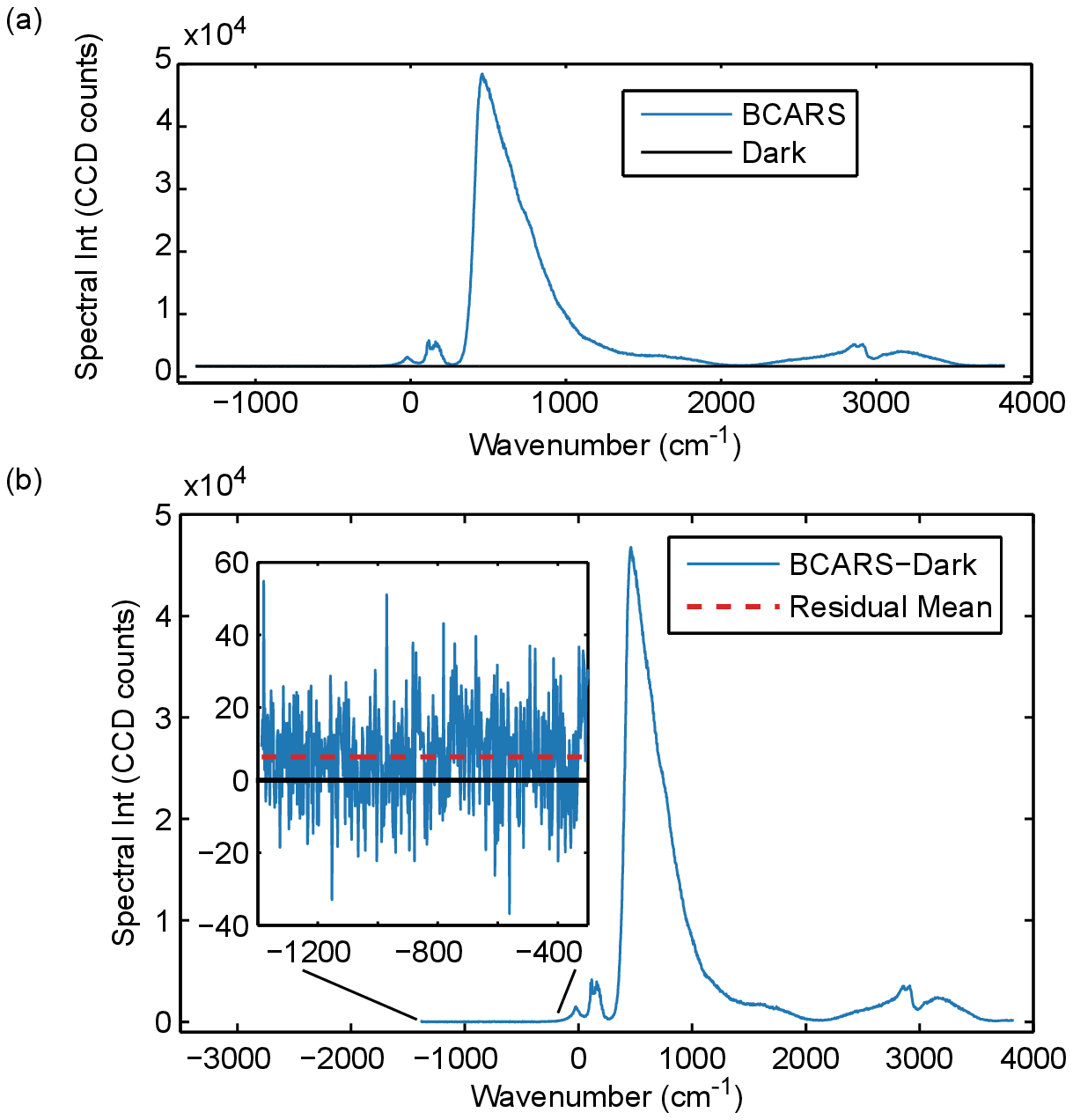}
\caption{Dark signal and residual subtraction. (a) BCARS and dark spectra. The acquired spectral window extends beyond the boundaries set by the optical filter set. (b) After dark signal subtraction, stray light or detector conditions may cause the baseline level to drift. This residual may be calculated from [ideally] unilluminated portions of the spectrometer and subtracted on a pixel-by-pixel basis.}
\label{WorkFlow:DarkSubtract}
\end{center}
\end{figure}

Typically, 100 to 1000 dark spectra are collected with no sample illumination, averaged, and subtracted from the BCARS and reference NRB spectra. Due to detector operating conditions or stray light, the exact level of dark signal may vary (slightly) from pixel-to-pixel or image-to-image. To remove this residual dark signal, we take advantage of the several hundred spectral pixels that ideally do not receive anti-Stokes photons. As shown in Figure \ref{WorkFlow:DarkSubtract} a, $\sim$350 spectral pixels ($\sim$900 cm$^{-1}$) are not illuminated, from which we can evaluate the mean (see Figure \ref{WorkFlow:DarkSubtract} b) and subtract. Although not explicitly discussed within this manuscript, deviations of the dark level will induce amplitude and phase errors within the extracted Raman-like spectra that are not straight-forward to remove.

\subsubsection{Denoising via the Anscombe Transformation and Singular Value Decomposition (SVD)}
SVD is a matrix factorization technique with many uses, including noise reduction, and has been applied previously to BCARS hyperspectral imagery\cite{CampJr2014,Masia2013,Chowdary2010,DiNapoli2014,Lee2011,Cicerone2012a}. BCARS hyperspectral imagery is unfolded into a two-dimensional matrix, $\mathbf{A}$, with rows representing the spectral axis and columns spatial content. The SVD algorithm factorizes this matrix into three components:
\begin{align}
\mathbf{A} = \mathbf{U}\mathbf{S}\mathbf{V^{\ast}},
\end{align}
where in this context, $\mathbf{U}$ contains the spectral bases (orthonormal eigenfunctions), $\mathbf{S}$ is a diagonal matrix containing the ``singular values" (SV) in descending order (descending average contribution), $\mathbf{V}$ describes the spatial distribution of the bases in $\mathbf{U}$, and `$\ast$' is the conjugate transpose. A simplified and typical method of de-noising is to analyze the normalized intensity of the singular values ($\mathbf{S}$), select a cut-off (C), set all higher diagonal elements $\mathbf{S}$ to 0, and to construct a denoised hyperspectral data matrix, $\mathbf{A}_{denoise}$, as:
\begin{align}
\mathbf{A}_{denoise} = \mathbf{U}\mathbf{S}\{1:C\}\mathbf{V^{\ast}}.
\end{align}
This method assumes that the signal and noise are separable and that (a) the signal is entirely enveloped in the lowest SVs and (b) the signal will be contained in consecutive SVs below a certain cut-off. Whether these conditions are met is determined by the signal-to-noise ratio of the signal and the noise distribution statistics. Of significant consequence, is that in its most general form, SVD (and the related principle component analysis [PCA]) assume the noise is additive and follows a normal distribution. In BCARS, and many other techniques, the noise of often of a mixed nature: containing (approximately) additive white Gaussian noise (AWGN) and Poisson noise. This is especially important in view of the recently-developed BCARS system, which generates spectra covering a large intensity range. This leads to mixed-noise with the mixing ratio varying dramatically across each individual spectrum. There is, therefore, a necessity to ``whiten" the noise as to be approximately constant (statistically) across each individual spectrum.

\begin{figure}[!ht]
\begin{center}
\includegraphics{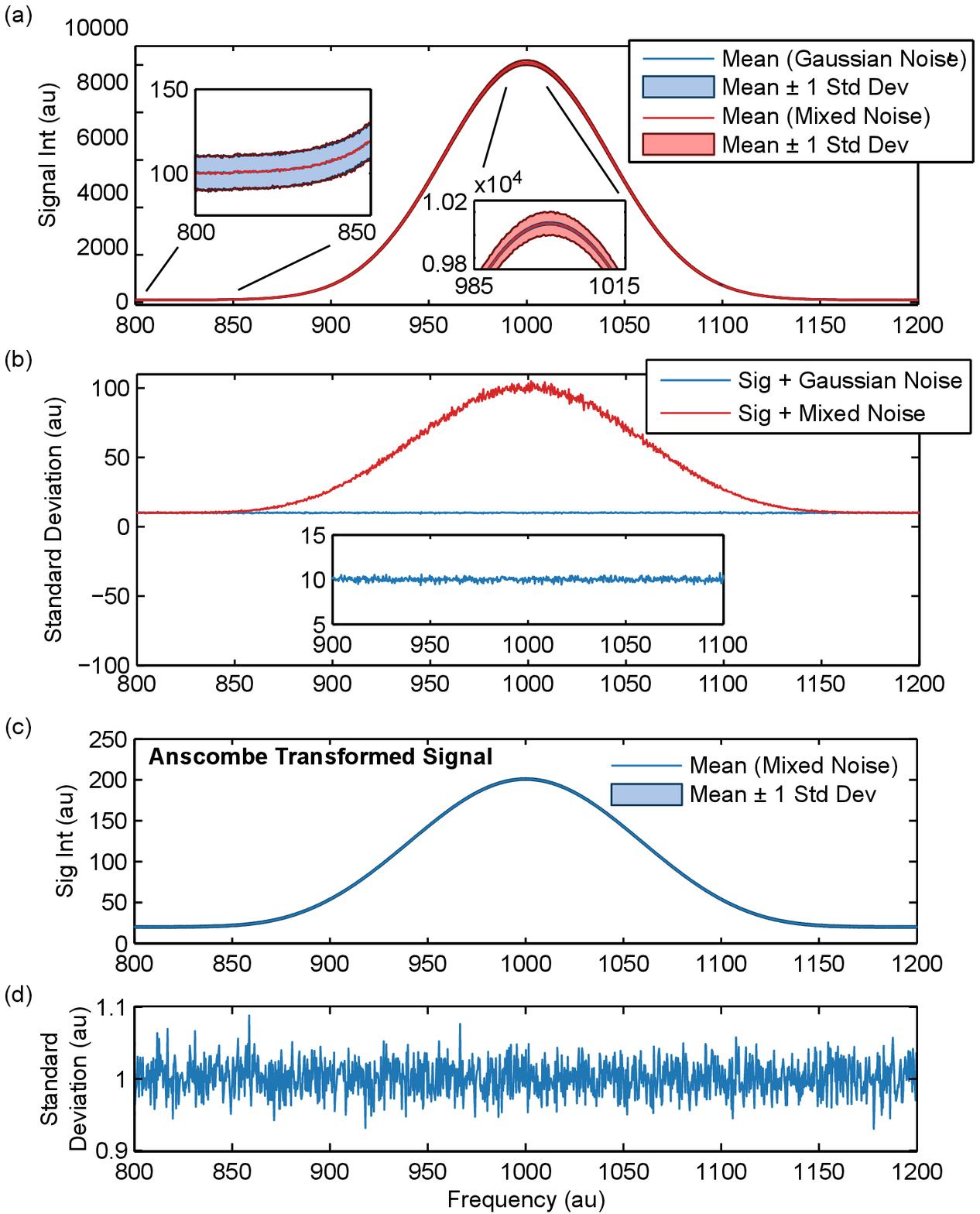}
\caption{Simulated variance stabilization via the Anscombe transformation. (a) Distribution of simulated signal over 1000 simulations with additive white Gaussian (AWGN) and mixed noise. (b) Standard deviation showing the intensity dependence of mixed noise. (c) Distribution of Anscombe transformed mixed noise signals. (d) Standard deviation showing normally distributed noise. }
\label{WorkFlow:SimAnscombe}
\end{center}
\end{figure}

One strategy is to perform ``variance-stabilization" using an Anscombe transformation\cite{Seymour1998,Makitalo2013}. Figure \ref{WorkFlow:SimAnscombe} a shows the results of 1000 simulations of a broad spectral peak containing AWGN (standard deviation, $\sigma$, of 10) or mixed noise. The level of the signal and the AWGN approximates the intensities found with actual BCARS experiments. Within spectral regions with weak intensity, the AWGN is the dominant noise source. At higher intensities, however, the Poisson noise is up to 10$\times$ larger, as shown by the plots detailing standard deviation in Figure \ref{WorkFlow:SimAnscombe} b. Applying the generalized Anscombe transformation to the 1000 simulated spectra and re-evaluating the standard deviation, as shown in Figure \ref{WorkFlow:SimAnscombe} c, there is no noticeable variation of the noise across the spectra.

\begin{figure}[!ht]
\begin{center}
\includegraphics{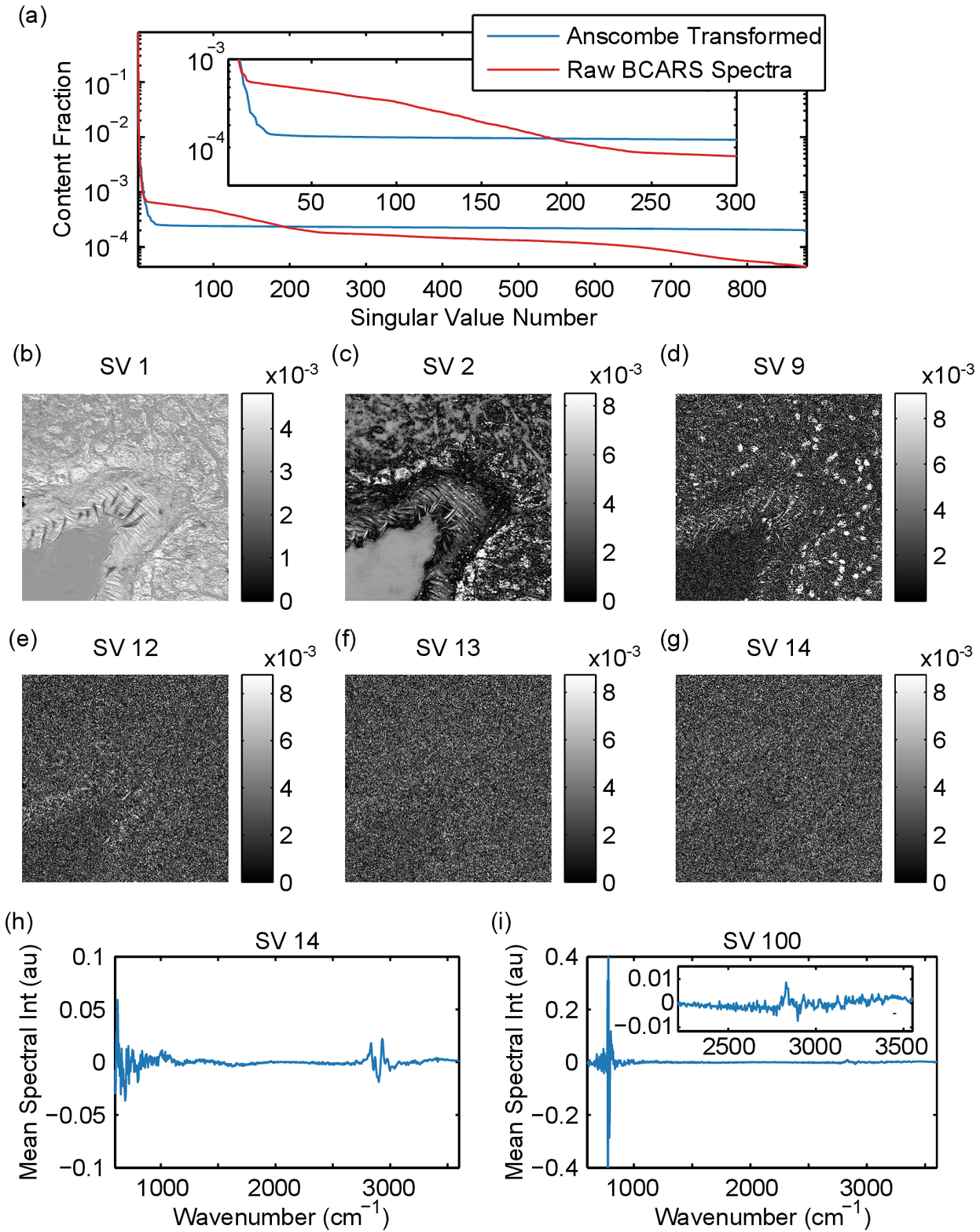}
\caption{Effect of mixed noise on SVD. (a) Content fraction of each basis spectrum with and without the Anscombe transformation. (b-g) Spatial distributions of selected SVs. (h) The 14th basis function shows spectral features and spectrally narrow noise (due to Poisson noise) though (g) does not indicate a contribution. (i) The 100th basis function still shows some spectral features and confined noise. With mixed noise characteristics of the measured spectra, SVD does not well-separate signal and noise components.}
\label{WorkFlow:SVD_No_Anscombe}
\end{center}
\end{figure}

Figure \ref{WorkFlow:SVD_No_Anscombe} a shows the normalized contribution of each spectral basis function (i.e., $S_{m,m}/\sum_m S_{m,m})$ during pre-processing of the murine pancreas tissue presented within the main text. As previously discussed, the typical approach to denoising is to set cutoff value based on singular value contribution (such as in Figure \ref{WorkFlow:SVD_No_Anscombe} a) or via qualitative analysis of the spatial and/or spectral components ($\mathbf{V}$ and $\mathbf{U}$, respectively). Figures S\ref{WorkFlow:SVD_No_Anscombe} b-g shows the spatial distribution of selected SVs (V-components). The spatial distributions of the lowest SVs clearly show features of the tissue, but by SV 13 or 14, it becomes less clear. Although there is no obvious spatial content of SV 14, the spectral basis function (Figure \ref{WorkFlow:SVD_No_Anscombe} h) shows Raman features. Additionally, the noise-like features below 1000 cm$^{-1}$ are due to the Poisson noise which is largest at the lowest wavenumbers in this particular BCARS system. Figure \ref{WorkFlow:SVD_No_Anscombe} i shows the 100th spectral basis function and, as expected, the Poisson noise has moved to lower wavenumber regions as this contribution is smaller. Additionally, one can see remaining Raman features near $\sim$2900 cm$^{-1}$. This supports the previous supposition that with the mixed-noise of BCARS spectra, SVD does not properly separate signal and noise, which also results in hundreds of SVs necessary to capture the signal accurately.

\begin{figure}[!ht]
\begin{center}
\includegraphics{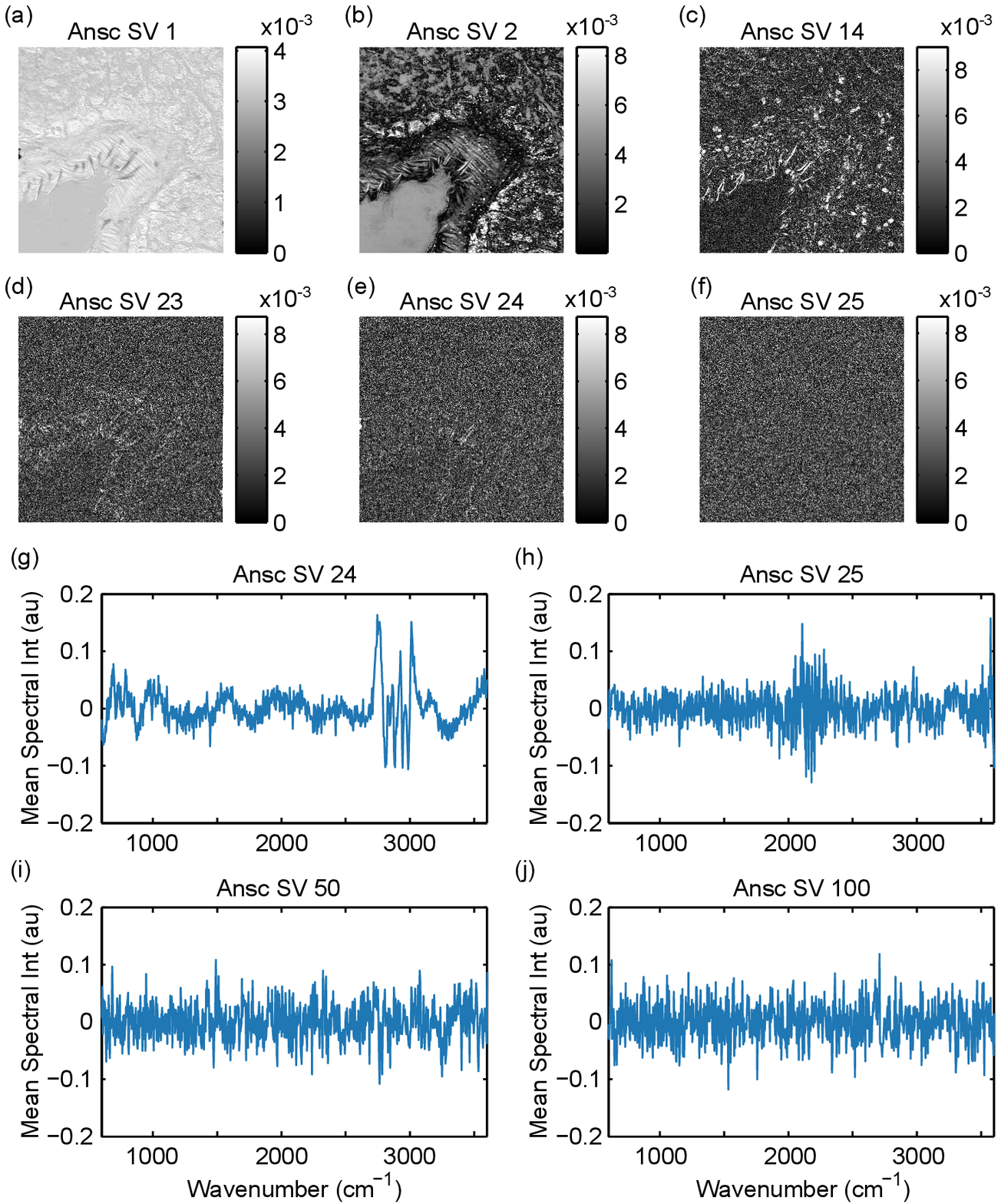}
\caption{SVD of Anscombe transformed spectra. (a-f) Spatial distribution of selected SVs (Ansc SV). (g) The 24th spectral basis function shows clear features as would be indicated by spatial content in (e). (h) The 25th SV appears to only contain noise (as does the corresponding (f)). By the 50th (i) and 100th (j) SVs, there is no noticeable remnants of signal. Without the Anscombe transformation, even the 100th SV (Figure \ref{WorkFlow:SVD_No_Anscombe} i) still contained Raman peak information.}
\label{WorkFlow:SVD_with_Anscombe}
\end{center}
\end{figure}

Figure \ref{WorkFlow:SVD_with_Anscombe} a-f shows the spatial distribution of SVs when using the Anscombe transformation on the BCARS spectra prior to SVD. In this case, SV 24 shows small spatial content and SV 25 none obvious. The spectral bases for these SVs (Figure \ref{WorkFlow:SVD_with_Anscombe} g,h) demonstrates that SV 24, which has spatial features, also contains spectral features, and SV 25 which has no spatial content also shows no obvious spectral content. Thus, there appears to be correlation between the spatial and spectral features. Additionally in both of these cases, there is no obvious noise variation across the spectrum (in contrast to Figure \ref{WorkFlow:SVD_No_Anscombe} h,i). Figure \ref{WorkFlow:SVD_with_Anscombe} i,j shows the spectral bases for the 50th and 100th SV. The non-variance stabilized SVD showed clear spectral features at the 100th SV, but using the Anscombe transformation, even the 50th SV appears to only contain noise.

Although this section qualitatively indicates that the Anscombe transformation assists the SVD in properly separating noise and signal, there is no clear indication that there exists a defined boundary between SVs containing predominantly signal and those predominantly noise. Additionally, the task of identifying SVs containing signal often involves human intervention and qualitative analysis. In the next section, we present a developed method of automating spectral and spatial feature analysis. As will be demonstrated, improper selection of SVs may result in distorted spectra, with improper Raman-peak amplitudes and even extra (or missing) Raman peaks.

\subsubsection{Automated Singular Value Selection}
\begin{figure}
\centering
\includegraphics{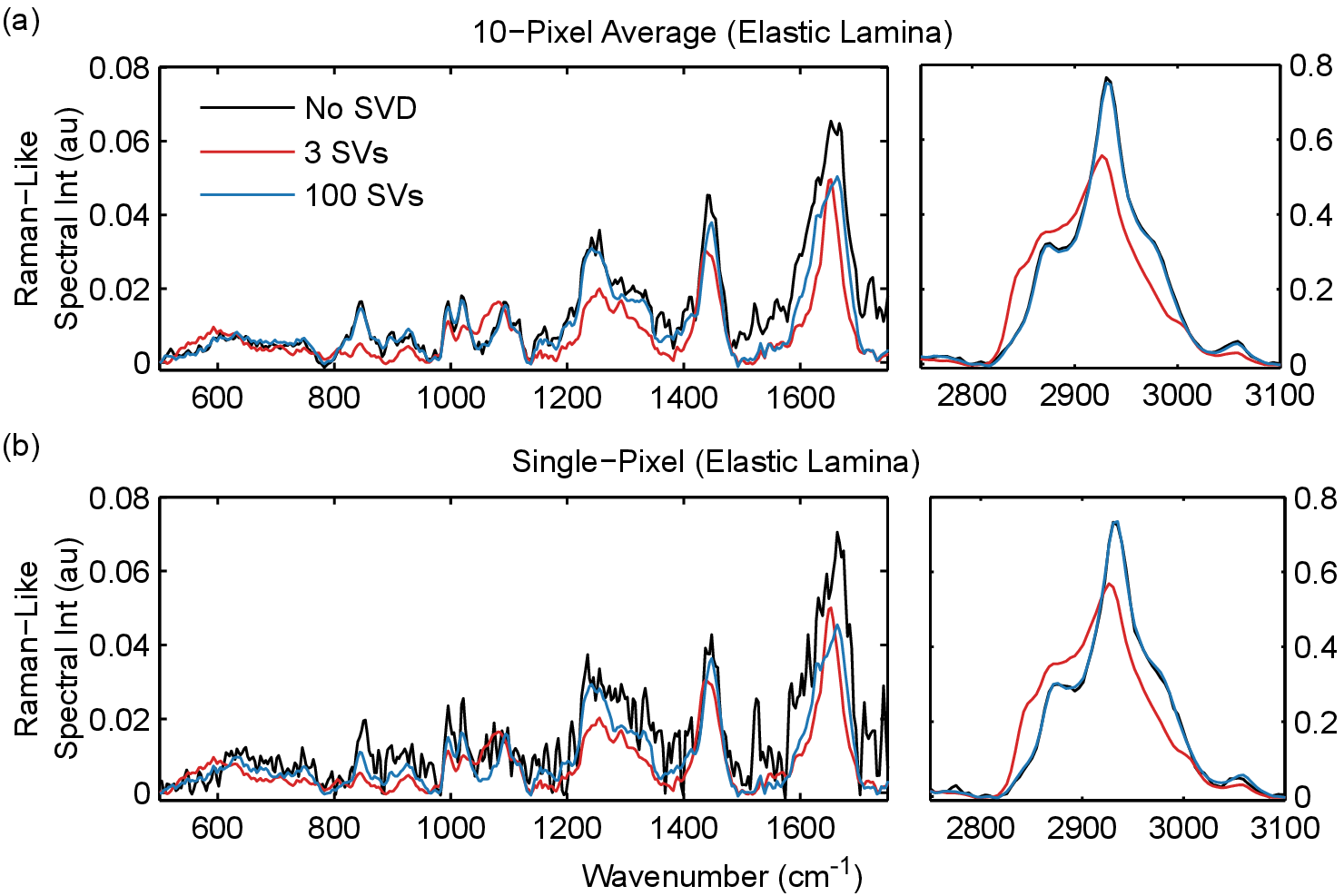}
\caption{Raman-like spectral distortions from improper singular value selection. (a) Mean spectra of 10 pixels within the elastic lamina of the murine pancreas tissue. Using only the first 3 SVs results in significant errors. Using the first 100 SVs shows significant improvement. (b) Single-pixel spectra from within the elastic lamina showing similar distortions with too few SVs incorporated.}
\label{WorkFlow:CompareSVD_Number_Distortion}
\end{figure}

Selecting the proper combination of SVs to reconstruct the original Raman content is of vital importance. Although the use of few SVs generates clear, attractive imagery, the underlying data may be distorted or completely erroneous. Figure \ref{WorkFlow:CompareSVD_Number_Distortion} a shows three spectra from within the internal elastic lamina of the artery (primarily composed of elastin) imaged within the main text of the manuscript. Each spectrum is the mean from the same 10 pixels but with varying number of SVs used in de-noising (or without SVD). The use of three SVs, shows significant spectral distortion. The spectral profile within the CH-stretch region would seem to indicate that the elastic lamina contains a significant lipid content (based on the pronounced 2850 cm$^{-1}$). Excluding baseline drift, the use of the first 100 SVs presents a significantly improved spectrum but with reduced efficacy of noise suppression (especially for a single pixel, as shown in Figure \ref{WorkFlow:CompareSVD_Number_Distortion} b). This indicates that there is a clear need to select enough SVs to maintain spectral integrity and few enough to provide significant noise reduction. The developed method uses both spectral and spatial features and the Fourier-domain noise statistics.

\begin{figure}[!ht]
\begin{center}
\includegraphics{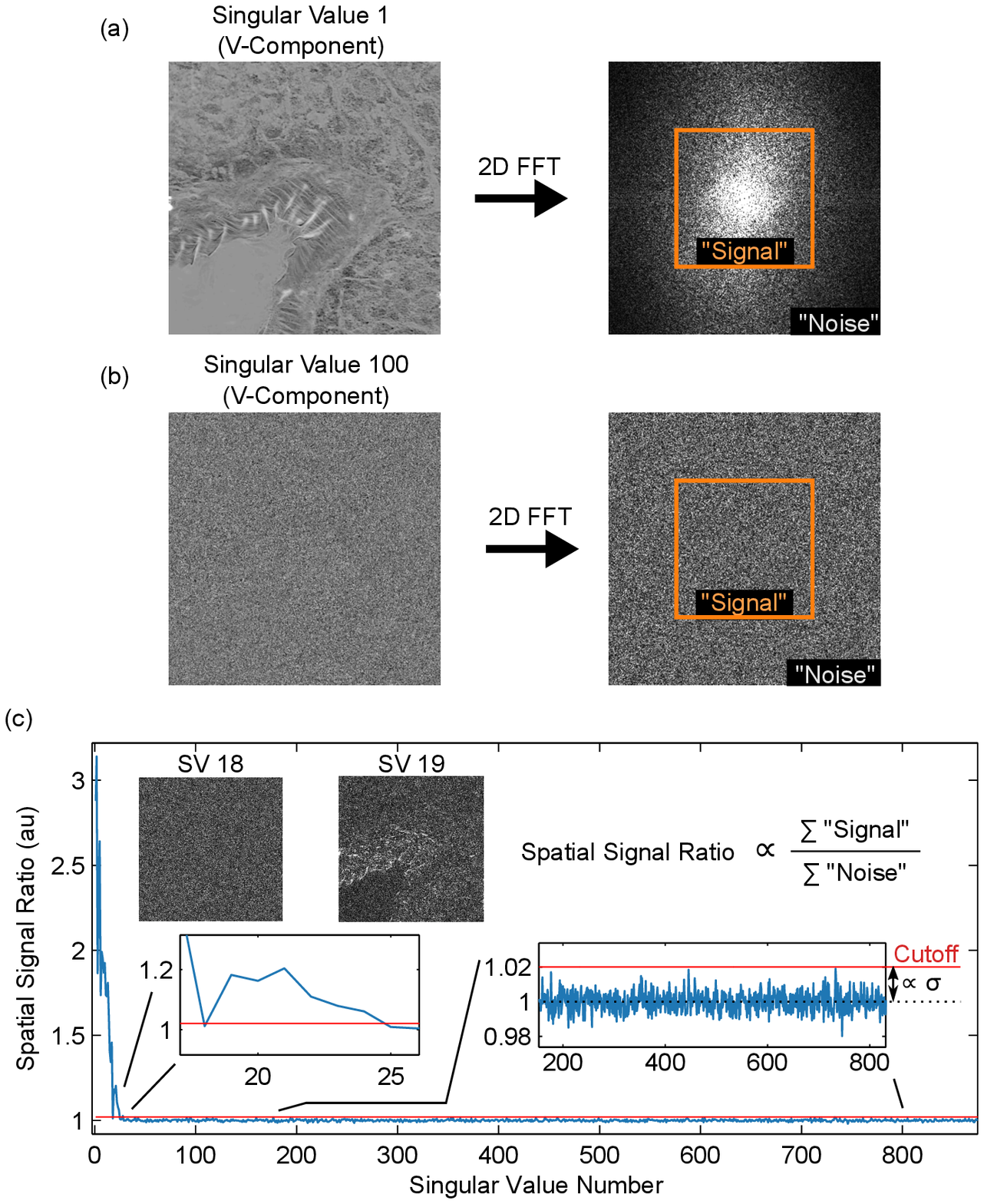}
\caption{Automated selection of singular values based on spatial content. (a,b) The spatial content is converted into the Fourier-domain from which ``signal" and "noise" regions are defined. (c) The spatial signal fraction is proportional to the total intensity ratio of the ``signal" and "noise" regions. A cut-off is selected based on a user-selected multiplier of the standard deviation of the spatial signal ratio at the highest SVs, which are assumed to predominantly contain noise.}
\label{WorkFlow:Auto_spatial}
\end{center}
\end{figure}

Figure \ref{WorkFlow:Auto_spatial} a,b demonstrates the concept of spatial domain SV selection. The absolute value of the spatial components (mean-subtracted) for each SV ($|\mathbf{V}|$) are transformed via a two-dimensional Fourier transform. A rectangular boundary is applied to separate the low-frequency components (assumed to contain primarily signal) and high-frequency components (assumed to be primarily noise). The ``spatial signal ratio" is defined as the sum of pixel intensities (complex modulus of the Fourier transform components) within the boundary divided by those outside the boundary (scaled by the number of pixels within and outside of the boundary). Figure \ref{WorkFlow:Auto_spatial} c plots the spatial signal ratio as a function of SV number. As expected the lowest SVs show the largest ratio. To develop a cut-off value, we assume that the highest SVs ($>$700) contain only noise. From the spatial signal ratio of these highest SVs, we calculate the standard deviation, $\sigma$. The cut-off value, for which all SVs below will be neglected, is heuristically determined as a multiple of $\sigma$. Based on visual inspection of $\mathbf{V}$, we select a multiplier of 3.5. This cut-off value may not be optimal, but it does appear, in our experience, fairly robust across images of different samples; although, we perform manual inspection of selected SVs prior to further progress in the pre-processing workflow. Figure \ref{WorkFlow:Auto_spatial} c shows this cut-off value in red. Between SV 15 and 20 (see Figure \ref{WorkFlow:Auto_spatial} c, inset), the spatial ratio drops below the cut-off (SV 18) then returns above (SV 19). Images of the spatial distribution of SV 18 and 19 (Figure \ref{WorkFlow:Auto_spatial} c) demonstrate that our method correctly categorized SV 18 as containing no obvious spatial components but SV 19 does. This demonstrates that there may be SVs with primarily noise components intermixed between SVs with primarily signal components.

\begin{figure}[!ht]
\begin{center}
\includegraphics{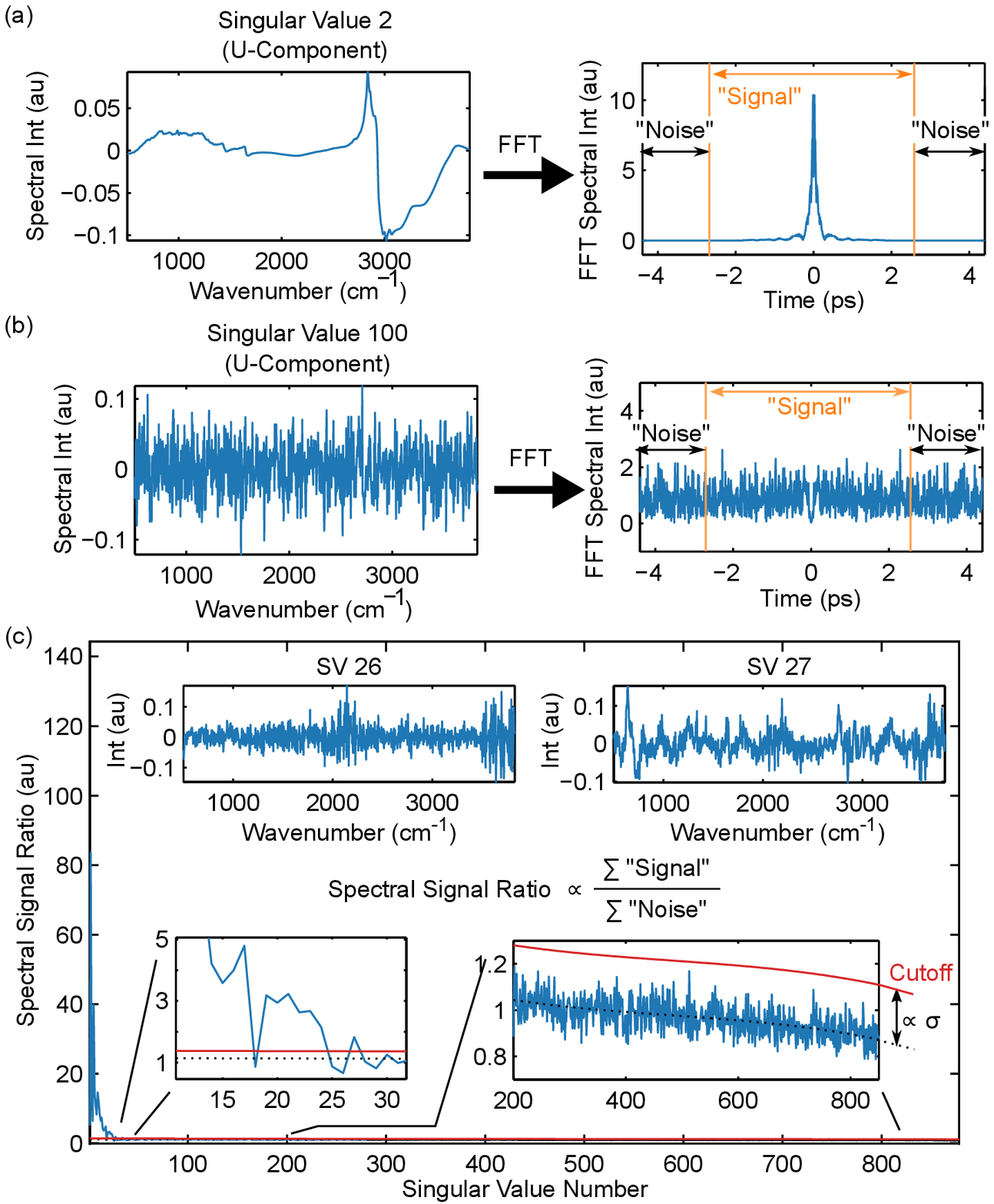}
\caption{Automated selection of SVs based on spectral content. (a,b) The spectral basis functions are converted into the time-domain from which ``signal" and "noise" regions are defined. (c) The spectral signal fraction is proportional to the ratio of the ``signal" and "noise" regions. A cut-off is selected based on a user-selected multiplier of the standard deviation of the spectral signal ratio at the highest SVs, which are assumed to predominantly contain noise.}
\label{WorkFlow:Auto_spectral}
\end{center}
\end{figure}

As illustrated in Figure \ref{WorkFlow:SVD_No_Anscombe} g,h, the lack of spatial components may or may not indicate a lack of spectral components. Thus, we have also developed an automated spectral basis set analysis technique. As pictographically described in Figure \ref{WorkFlow:Auto_spectral} a,b, the spectral component selection tool analyzes the statistics of the $\mathbf{U}$ components within the Fourier domain (time-domain). Again, a ratio is determined between the sum of the signal contribution within a low-frequency window and outside this window (scaled to number of pixels). The half-width of this window is set to correspond to the temporal duration of our probe source ($\sim$ 3.4 ps). Again, we assume that the highest SV components contain primarily noise and calculate the standard deviation. The spectral components, however, show a mean-line drift, which we fit with a low-order polynomial (in this case, third-order). The cut-off is a multiple (4, in this case) of the standard deviations from this mean. Figure \ref{WorkFlow:Auto_spectral} c (inset) demonstrates qualitatively correct identification of SV 26 with no noticeable Raman components and SV 27 with signal features.

Table S1 shows the time of each pre-processing step in automatically processing the murine pancreas tissue presented within the main text. The computer was a Dell Optiplex 9010 with quad-core Intel i7-3770 CPU at 3.4 GHz, with 16 GB of memory, and running MATLAB R2013a. The total processing time was approximately 28 minutes. At 95\% of the computation time, the most intensive process was the automated detrending using asymmetric least squares (ALS). This method, although robust and reuqiring no user intervention, requires multiple matrix inversions, which do not facilitate processing several spectra in parallel. It should be noted that the ALS algorithm, as developed, uses the CHOLMOD implementation of fast sparse matrix inversion for optimal performance. Future work will investigate alternative means of automated detrending. Excluding the ALS step, the total processing time was approximately 90 seconds ($<$ 1 ms/spectrum).

\begin{table}[h]
\begin{center}
\begin{tabular}{|c|c|c|}
\hline  Sub-Process & Total Computation Time (s) & Time/Pixel (ms) \\ 
\hline  Dark Subtraction & 0.74 & 0.01 \\ 
\hline  Residual Baseline Subtraction & 1.14 & 0.01 \\ 
\hline  Anscombe Transform & 0.48 & 0.01 \\ 
\hline  SVD & 41.71 & 0.46\\ 
\hline  Automated SV Selection & 3.67 & 0.04\\ 
\hline  Inverse Anscombe Transform & 5.18 & 0.06\\ 
\hline  Phase Retrieval (KK) & 17.66 & 0.20\\ 
\hline  Phase Detrending (ALS) & 1564.91 & 17.44\\ 
\hline  Spectral Scaling &  19.69 & 0.22\\ 
\hline \textbf{Total} & \textbf{1655.18	} & \textbf{18.45}\\
\hline 
\end{tabular} 
\caption{Computation time of automated pre-processing of hyperspectral BCARS image of murine pancreas. The total processing time was $\sim$28 minutes.}
\end{center}
\end{table}

\begin{figure}[!ht]
\begin{center}
\includegraphics{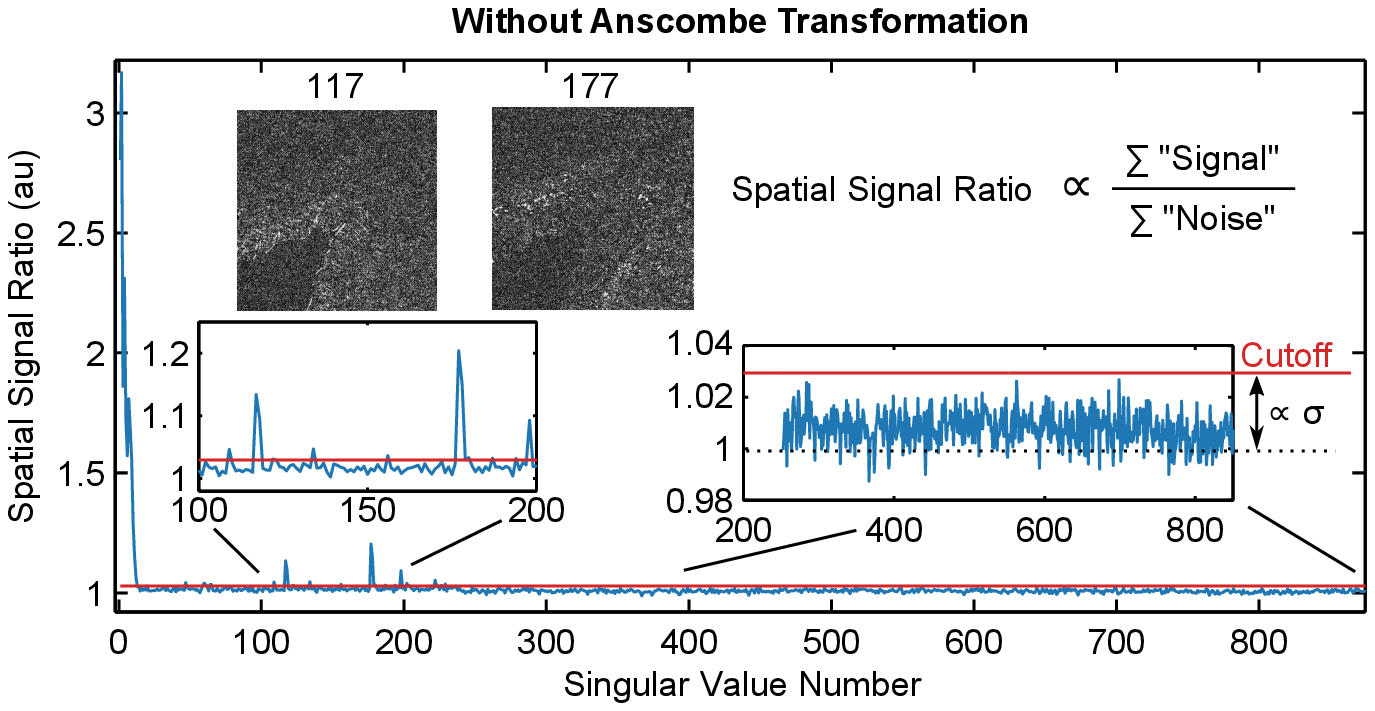}
\caption{Automated selection of singular values based on spatial content without use of the Anscombe transformation. In comparison to the results in Figure \ref{WorkFlow:Auto_spatial}, without the use of the Anscombe transformation, SVs showing clear spatial contributions are spread over several hundreds SVs, such as 117 and 177 (inset). This automated method improves SV-selection time as it would be laborious for a user to manually inspect hundreds of SVs. Additionally, the automated method removes many of the SVs that contain primarily noise; although, they are at lower SVs.}
\label{WorkFlow:Auto_spatial_NO_Ansc}
\end{center}
\end{figure}

\begin{figure}[!ht]
\begin{center}
\includegraphics{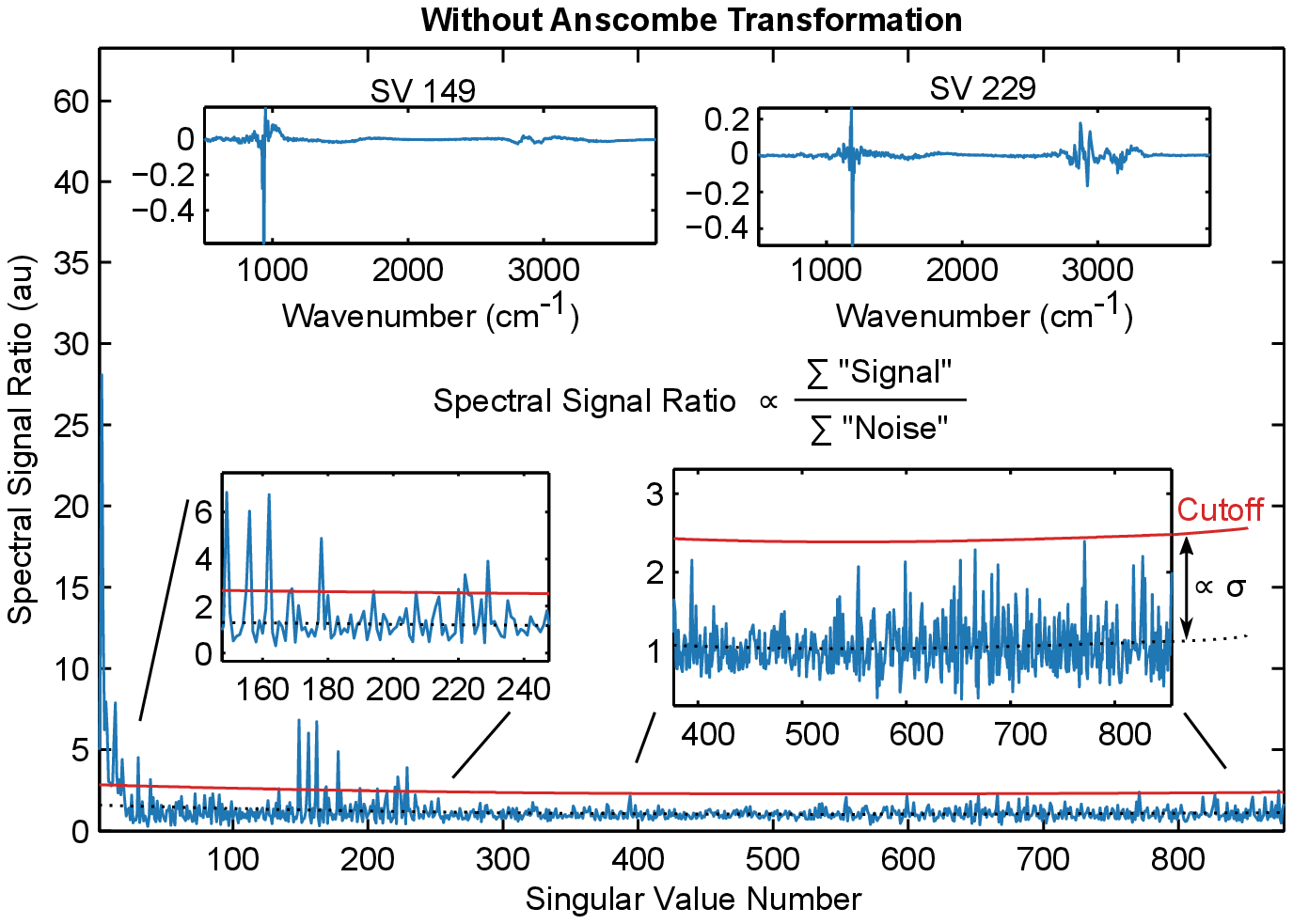}
\caption{Automated selection of singular values based on spectral content without use of the Anscombe transformation. In comparison to the results in Figure \ref{WorkFlow:Auto_spectral}, without the use of the Anscombe transformation, SVs showing clear spectral contributions are spread over several hundreds SVs, such as 149 and 229 (inset). Additionally, without the Anscombe transformation, there is an obvious mixing between signal contributions and those due to Poisson noise.}
\label{WorkFlow:Auto_spectral_NO_Ansc}
\end{center}
\end{figure}

For comparison with the demonstrations above, Figures S\ref{WorkFlow:Auto_spatial_NO_Ansc} and S\ref{WorkFlow:Auto_spectral_NO_Ansc} show the automated spatial and spectral analysis results for non-Anscombe transformed BCARS data.

\section{Results}
\subsection{The Maximum Entropy Method (MEM) and Phase Error Correction Applied to Simulated Spectra}
\begin{figure}[!ht]
\begin{center}
\includegraphics{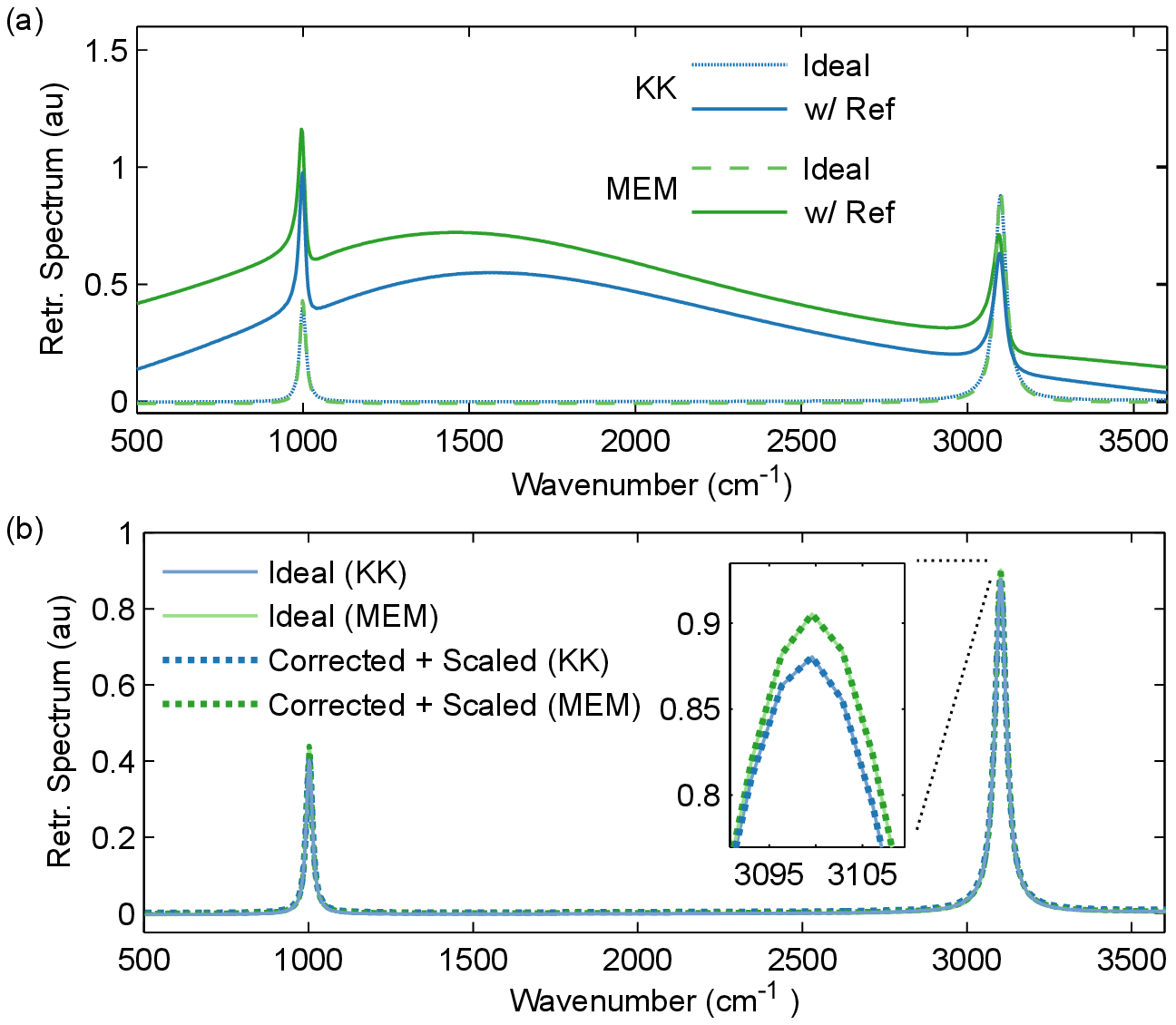}
\caption{Phase detrending and scaling applied to the maximum entropy method (MEM). (a) Comparison of extraction of Raman signatures using the KK and MEM methods with a known NRB (``ideal") and with a reference NRB (``Ref"). Both methods produce distorted spectra due to phase and amplitude errors when using a reference NRB. (b) Applying the previous described phase error corrections technique produces respective spectra for the MEM and KK that are identical whether using the actual NRB or a surrogate reference. Differences between the MEM and KK are due to the underlying numerical algorithms and not explicitly to due to phase or amplitude errors.}
\label{KK_MEM}
\end{center}
\end{figure}

The MEM performs phase-retrieval based on information theory ground\cite{Vartiainen1992}. Previously, the MEM and KK methods were demonstrated to be ``functionally" equivalent\cite{Cicerone2012}. To that end, we performed phase retrieval via the MEM on simulated BCARS spectra containing two peaks under ideal (known NRB) conditions and with a surrogate reference NRB. As shown in Figure \ref{KK_MEM} a, the retrieved Raman spectra differ between the MEM and KK methods, with the MEM showing a slightly larger baseline drift. Upon phase detrending and scaling, the Raman-like spectra extracted using reference spectra agree with those with the known NRB. Differences in peak amplitudes between the MEM and KK ($<$5\%) are due to the underlying algorithmic differences between the KK and MEM methods. Comparison of these methods is presented in further detail in Ref. \citenum{Cicerone2012}.

\subsection{BCARS Spectra of Glycerol}
\begin{figure}
\centering
\includegraphics{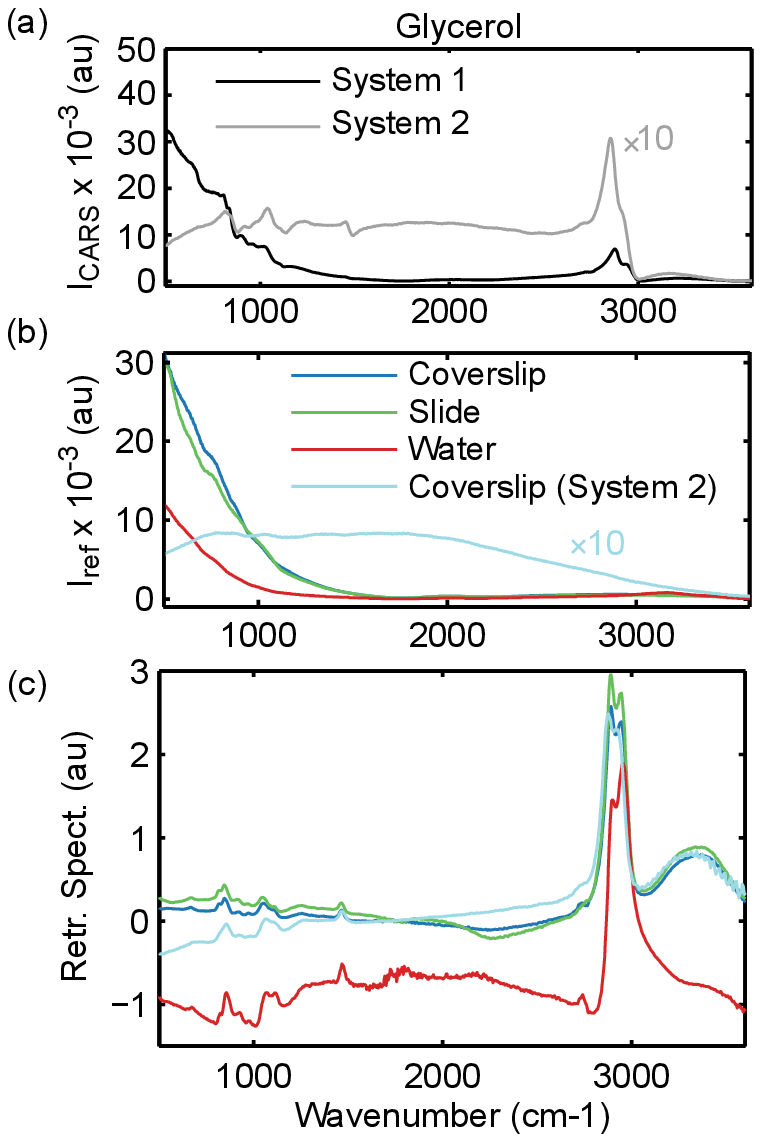}
\caption{Extraction of Raman-like spectra of glycerol using difference NRB references and difference BCARS platforms. (a) Raw BCARS spectra collected ona  recently-developed BCARS platform (``System 1") and a more traditional implementation (``System 2"), showing significantly difference system responses. (b) Raw BCARS spectra of 3 common NRB reference materials: glass coverslip, glass microscope slide, and water. Additionally, a raw spectrum from glass coverslip was collected on System 2. (c) Raman-like spectra extracted from (a) using spectra in (b) without amplitude or phase error correction, showing different shapes and amplitudes for corresponding peaks.}
\label{CompareGlycerolSpectra_Systems_150609}
\end{figure}

As described within the main text, glycerol spectra were collected on two separate BCARS platforms and pre-processed with different reference NRB spectra. Figure \ref{CompareGlycerolSpectra_Systems_150609} a shows BCARS spectra of glycerol collected on a recently-developed platform\cite{CampJr2014} (``System 1") and an older system\cite{Parekh2010,Lee2011} (``System 2"). There is a clear difference in the system responses with the newer system showing marked enhancement at the lowest wavenumbers. Figure \ref{CompareGlycerolSpectra_Systems_150609} b shows the BCARS spectra of 3 different reference materials (and coverslip glass acquired on System 2). These spectra show differences in amplitude and shape, which will be discussed in detail in the next subsection. Figure \ref{CompareGlycerolSpectra_Systems_150609} c shows the retrieved Raman spectra using the KK relation (no error correction performed). Inspection of the Raman peaks (such as the CH-stretch peaks) indicates there is substantial differences in retrieved amplitude and phase.

\subsection{Time-Window Self-Referencing to Directly Capture the Approximate NRB of the Sample and the Raman Peaks of Reference Materials}
Time-window self-referencing (TWSR) is a method we have implemented to capture a CARS spectrum that predominantly contains the NRB with reduced contributions of the Raman vibrational components. This techniques can be used to retrieve an approximate NRB spectrum on a pixel-by-pixel basis, requiring a second image to be acquired, but this falls outside of the scope of this manuscript. Rather we will use TWSR to enable extraction of the Raman signature of typical NRB reference materials to highlight that these materials do, in fact, contain Raman peaks that will affect the final spectral retrieval.

\begin{figure}[!ht]
\begin{center}
\includegraphics{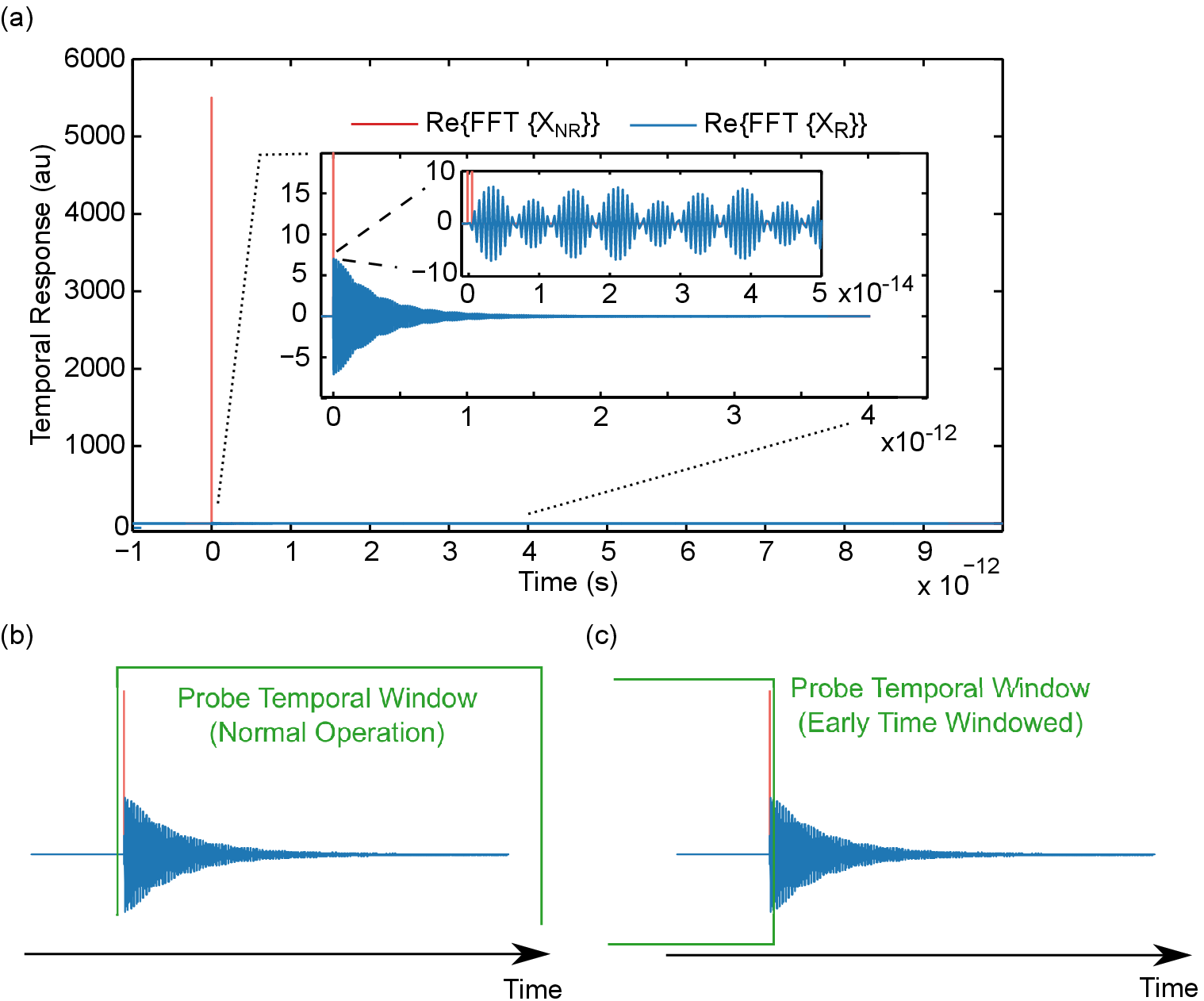}
\caption{Time-window self-referencing (TWSR) to retrieve the approximate NRB of the sample. (a) Temporal response of the material calculated from the fast Fourier transform (FFT) of the nonlinear susceptibility. (b) Under normal operation, the probe source delay (multiplicative in the time-domain) is set to acquire the entire temporal evolution of the Raman/CARS signal. (c) One can acquire a spectrum that is predominantly NRB by setting the probe delay to an early time point. There still exists a Raman contribution, but it is weaker and the peaks are significantly broadened owing the effective, limited spectral resolution.}
\label{TimeWindowing}
\end{center}
\end{figure}

As described within the main text, the ``effective" nonlinear susceptibility, $\widetilde{\chi}$, is related to the nonlinear susceptibility as:
\begin{align}
\widetilde{\chi}(\omega) = \chi(\omega)\ast E_{pr}(\omega),
\end{align}
where $\ast$ is the convolution operation and $E_{pr}$ is the electric field of the probe source. In essence the spectral resolution of the CARS spectrum is determined by the spectral narrowness of the probe source\cite{Cheng2004}: the spectrally narrower the probe source, the higher the resolution of the CARS spectrum. From a time-domain perspective, this can be viewed as how much temporal information is acquired. One can describe the effective time response of the nonlinear susceptibility, $\widetilde{R}(t)$, as:
\begin{align}
\widetilde{R}(t) = \mathcal{F}^{-1}\left \lbrace \widetilde{\chi}^{(3)}(\omega) \right \rbrace = \underbrace{\mathcal{F}^{-1}\left \lbrace \chi^{(3)}(\omega) \right \rbrace}_{R(t)} \underbrace{\mathcal{F}^{-1}\left \lbrace E_{pr}(\omega) \right \rbrace}_{E_{pr}(t)},
\end{align}
where $\mathcal{F}^{-1}$ is the inverse Fourier transform, $R(t)$ is the time response of the material nonlinear susceptibility, and $E_{pr}(t)$ is the temporal field profile of the probe source. Figure \ref{TimeWindowing} a shows the time response of a simulated nonlinear susceptibility (resonant and nonresonant contributions) containing two Raman peaks and a broad (25,000 cm$^{-1}$ full-width half-max) nonresonant background (see Figure \ref{NonlinearSusceptWindowEdge} a for plots of the individual terms in the frequency-domain). Notice that the nonresonant term only contributes a signal for a brief duration (femtoseconds) but the Raman contributions decays over several picoseconds. Under normal spectroscopic collection, as described in Figure \ref{TimeWindowing} b, the temporal overlap of the material stimulation (via pump and Stokes sources) and the probe source is set to maximize the total energy collection (i.e., from time 0, on); thus, maximizing the temporal duration captured and the spectral resolution. In the TWSR technique, the probe source is temporally offset as to only capture the earliest moments of signal creation (Figure \ref{TimeWindowing} c); thus, acting as a femtosecond probe. The signal generation from the nonresonant component of the nonlinear susceptibility is the same as under normal operating conditions, but little of the Raman decay is acquired. The ``early time" (ET) spectrum will serve as an approximate NRB measurement.

\begin{figure}[!ht]
\begin{center}
\includegraphics{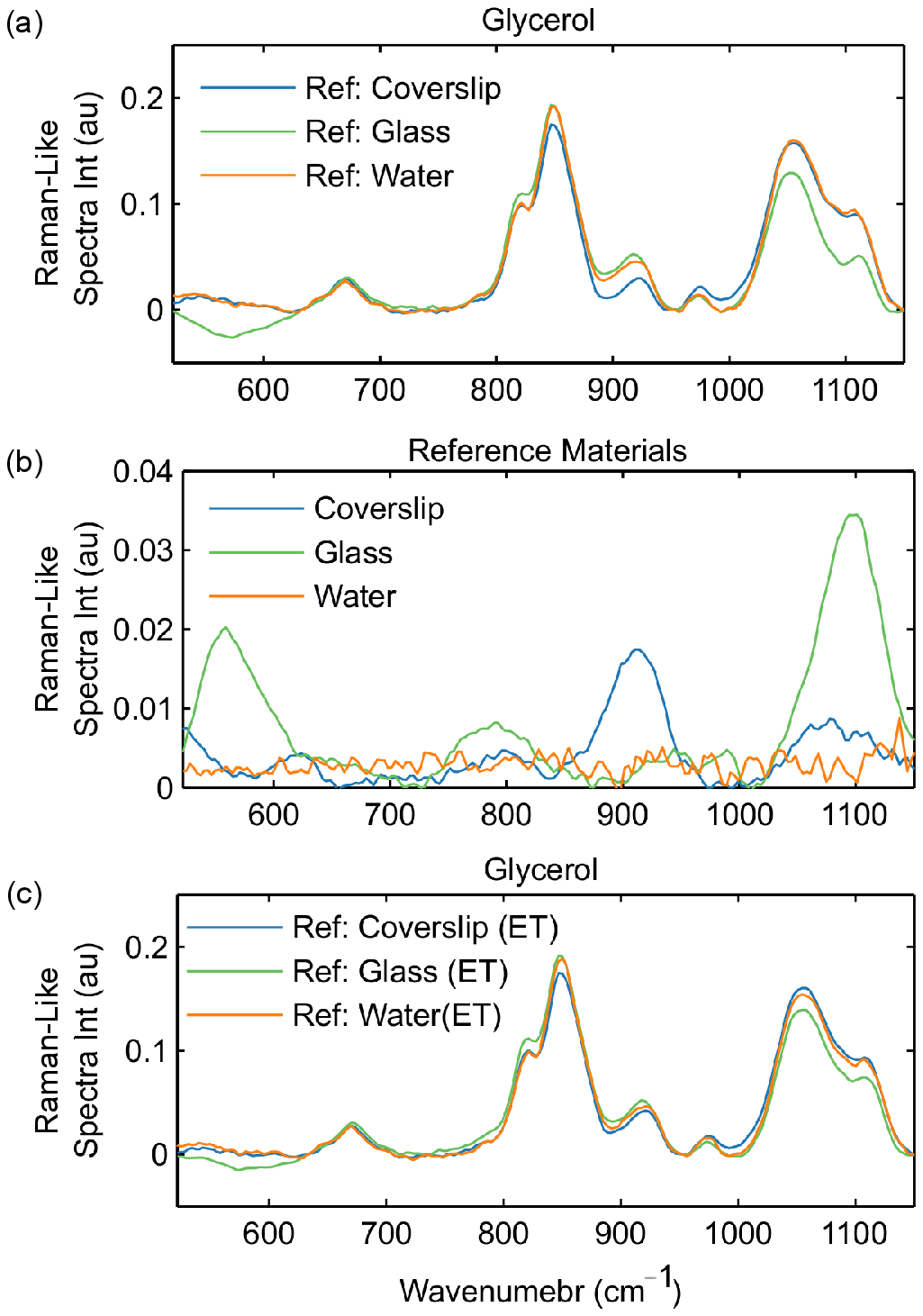}
\caption{Using temporal dynamics of the CARS generation process to improve the utility of NRB reference measurements. (a) The retrieved Raman-like spectra of glycerol (with the use of reference NRB measurements and corrected via phase detrending and scaling) showing close agreement but some deviations. (b) Raman-like spectra extracted from common reference NRB materials using TWSR to collect an approximate NRB spectrum of each material. The peaks in (b) correlate with the differences between spectra in (a). (c) Extracted Raman-like glycerol spectra using NRB reference spectra that were collected under early-time conditions to reduce the Raman contributions. Comparing (a) and (c), the spectral differences are reduced in amplitude by $\sim$2x.}
\label{GlycerolEarlyLateTime}
\end{center}
\end{figure}

As an experimental demonstration of the effect of Raman peaks within common reference materials, Figure \ref{GlycerolEarlyLateTime} a shows the retrieved Raman-like spectra of glycerol within the lower wavenumber region using reference NRB (normal temporal probe settings). The spectra show distinct similarity, but there are certain regions of deviation at $\sim$560 cm$^{-1}$, $\sim$920 cm$^{-1}$, and $\sim$1100 cm$^{-1}$. By using the spectra gathered from these reference materials and the TWSR technique to capture an approximate NRB spectrum, one can retrieve Raman-like spectra from the reference materials themselves as shown in Figure \ref{GlycerolEarlyLateTime} b. Comparing Figure \ref{GlycerolEarlyLateTime} (a,b), it is apparent that the Raman peaks of coverslip and glass slide correspond to the same spectral regions of deviation. Figure \ref{GlycerolEarlyLateTime} c shows the retrieved Raman-like spectra of glycerol using TWSR with the reference materials, reducing their Raman spectral contribution. The use of TWSR in this manner reduces the deviation of spectral components by a factor of $\sim$2.

\subsection{Tissue Imaging and Histogram Analysis}
The pseudocolor imagery and analysis was performed to highlight general protein content, smooth muscle (predominantly actin/myosin), and DNA/RNA. For protein 2937 cm$^{-1}$ - 2882 cm$^{-1}$ was used, which effectively suppresses the lipid content. For smooth muscle, a peak at 1339 cm$^{-1}$ was used. As this particular peak is a portion of a shoulder that is spectrally broad, containing several neighboring peaks, a linear interpolant was calculated between 1288 cm$^{-1}$ and 1360 cm$^{-1}$, subtracted, and the peak amplitude at 1339 cm$^{-1}$ was calculated. For DNA/RNA, the peak at 785 cm$^{-1}$ was used. To mitigate the effect of residual baseline, this peak amplitude was calculate relative to a linear interpolant calculated between the neighboring troughs at 763 cm$^{-1}$ and 820 cm$^{-1}$.

\begin{figure}[!ht]
\begin{center}
\includegraphics{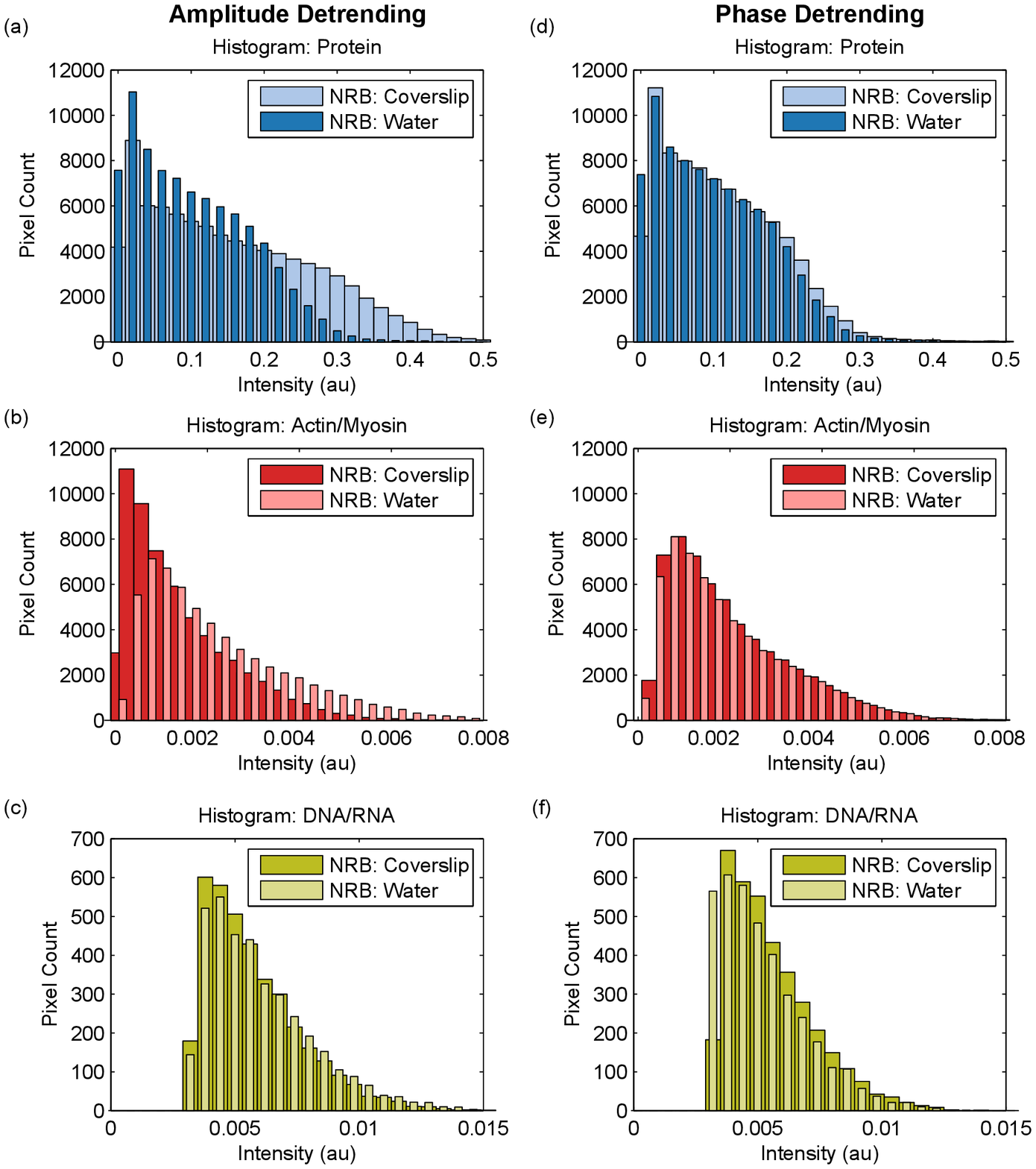}
\caption{Histogram analysis of spectral peaks amplitudes using traditional amplitude detrending and the new phase detrending methods. (a-c) Pixel-by-pixel comparison of Raman-like peak intensities using different NRB surrogate materials and traditional amplitude detrending. (d-f) Pixel-by-pixel comparison of Raman-like peak intensities using different NRB surrogate materials and the developed phase detrending and scaling method. Note: histogram bins are always the same width-- bar widths adjusted for visual clarity.}
\label{MouseHistogram}
\end{center}
\end{figure}

As described in the main text, phase detrending and scaling can significantly reduce the effect of using reference NRB spectra. Figure \ref{MouseHistogram} a-c shows histograms of the calculated peak amplitudes relating to protein, smooth muscle, and DNA/RNA as a function of reference NRB material corrected with traditional amplitude detrending. For the case of general protein (Figure \ref{MouseHistogram} a), the use of glass coverslip as the surrogate NRB retrieves peak amplitudes with a larger mean and broader distribution than with water. The converse is true for smooth muscle (actin/myosin) for which retrieval increases the amplitude distribution to higher values (Figure \ref{MouseHistogram} a). For DNA/RNA, the retrieved amplitudes are relatively similar. Figure \ref{MouseHistogram} d-f demonstrates that the use of phase detrending and scaling significantly improves the similarity of retrieved amplitudes regardless of reference material used.

\section*{Disclaimer}
Any mention of commercial products or services is for experimental clarity and does not signify an endorsement or recommendation by the National Institute of Standards and Technology. The authors declare no competing financial interests.
\clearpage

\bibliography{QuantBCARS}
\end{document}